\documentclass[12pt,peerreview]{IEEEtran}
\usepackage{amsmath,amssymb}
\usepackage{graphicx}
\usepackage[dvipsnames]{xcolor}
\usepackage[T1]{fontenc}
\usepackage{hyperref}
\usepackage[noadjust]{cite}
\usepackage{subcaption}
\usepackage{booktabs}
\usepackage{multirow}
\usepackage{tikz}
\usetikzlibrary{arrows.meta, positioning, shapes.geometric, calc}
\usepackage{pgfplots}
\pgfplotsset{compat=1.18}
\usepackage{xspace}

\newcommand{\hawaii}{Hawaii}

\definecolor{blue}{HTML}{0072BD}
\definecolor{red}{HTML}{D95319}
\definecolor{yellow}{HTML}{EDB120}
\definecolor{purple}{HTML}{7E2F8E}
\definecolor{green}{HTML}{77AC30}
\definecolor{brown}{HTML}{8C564B}
\definecolor{pink}{HTML}{E377C2}
\definecolor{black}{HTML}{000000}
\newcommand{\blue}{\textcolor{blue}{\texttt{Blue}}\xspace}
\newcommand{\red}{\textcolor{red}{\texttt{Red}}\xspace}
\newcommand{\yellow}{\textcolor{yellow}{\texttt{Yellow}}\xspace}
\newcommand{\purple}{\textcolor{purple}{\texttt{Purple}}\xspace}
\newcommand{\green}{\textcolor{green}{\texttt{Green}}\xspace}
\newcommand{\brown}{\textcolor{brown}{\texttt{Brown}}\xspace}
\newcommand{\pink}{\textcolor{pink}{\texttt{Pink}}\xspace}
\newcommand{\black}{\textcolor{black}{\texttt{Black}}\xspace}

\usepackage{acronym}
\newacro{itu}[ITU]{International Telecommunication Union}
\newacro{ieee}[IEEE]{Institute of Electrical and Electronics Engineers}
\newacro{pn}[PN]{pseudo-noise}
\newacro{mmse}[MMSE]{minimum mean squared error}
\newacro{dll}[DLL]{delay-locked loop}
\newacro{pll}[PLL]{phase-locked loop}
\newacro{bpsk}[BPSK]{binary phase-shift keying}
\newacro{qpsk}[QPSK]{quadrature phase-shift keying}
\newacro{mace}[MACE'10]{2010 Mobile Acoustic Communications Experiment}
\newacro{space}[SPACE'08]{2008 Surface Processes and Acoustic Communications Experiment}
\newacro{kam}[KAM'11]{2011 Kauai Acomms Multidisciplinary University Research Initiative}
\newacro{jamstec}[JAMSTEC]{Japan Agency for Marine-Earth Science and Technology}
\newacro{lms}[LMS]{least mean squares}
\newacro{rls}[RLS]{recursive least squares}
\newacro{snr}[SNR]{signal-to-noise ratio}
\newacro{dfe}[DFE]{decision-feedback equalizer}
\newacro{atoc}[ATOC]{Acoustic Thermometry of Ocean Climate}

\begin{document}

\title{Underwater Acoustic Channel Library}

\author{\IEEEauthorblockN{Zhengnan~Li,~\IEEEmembership{Member,~IEEE,}
Mandar~Chitre,~\IEEEmembership{Fellow,~IEEE,}
Diego~A.~Cuji,~\IEEEmembership{Member,~IEEE,}
James~Preisig,~\IEEEmembership{Fellow,~IEEE,} 
Andrew~C.~Singer,~\IEEEmembership{Fellow,~IEEE,}
Milica~Stojanovic,~\IEEEmembership{Fellow,~IEEE,}
and~Paul~van~Walree,~\IEEEmembership{Member,~IEEE}}

\thanks{Z. Li is with the Department of Electrical and Computer Engineering, Northeastern University, Boston, MA, USA, and the Department of Electrical and Computer Engineering, The University of Alabama, Tuscaloosa, AL, USA (e-mail: zhengnan.li@ua.edu).}
\thanks{M. Chitre is with the Department of Electrical and Computer Engineering, and the ARL, Tropical Marine Science Institute of the National University of Singapore.}
\thanks{D. A. Cuji is with the Department of Electrical and Computer Engineering, Stony Brook University, Stony Brook, NY, USA.}
\thanks{J. Preisig is with JPAnalytics.}
\thanks{A. C. Singer is with the Department of Electrical and Computer Engineering, Stony Brook University, Stony Brook, NY, USA.}
\thanks{M. Stojanovic is with the Department of Electrical and Computer Engineering, Northeastern University, Boston, MA, USA.}
\thanks{P. van Walree is with the Norwegian Defence Research Establishment (FFI), Horten, Norway.}
}

\maketitle

\begin{abstract}
    The development of communication systems critically depends on realistic channel models, yet there are no widely accepted standards in the underwater acoustic communications community. The situation is in stark contrast to terrestrial radio communications, where  channel models have been standardized and are widely available. To address this gap, we present an open-access library of underwater acoustic channels derived from field experiments conducted across geographically distinct locations and varying propagation conditions, including shallow and deep water, short and long range, and fixed and mobile platforms. Each channel is described by a time-varying impulse response extracted from at-sea recordings using an adaptive algorithm that separately identifies the multipath structure and  the Doppler-induced phase and delay drift. Each channel is also accompanied by a site-specific ambient noise model, which captures the statistics of  colored Gaussian noise and  impulsive noise. Spatial diversity reception across an array of hydrophones is supported for most channels, while time diversity  is included for single-hydrophone scenarios. The library is accompanied by a simple replay interface through which  a user supplies an arbitrary transmit signal, passes it through a chosen channel, adds noise at a desired signal-to-noise ratio (SNR), and obtains the received signal. The models are validated by comparing the output SNR of a baseline receiver operating on replayed signals with its performance on the original at-sea recordings, demonstrating close agreement across all channels. The library, including all channel impulse responses, noise model parameters, and replay software, is freely available for download as an open-source package.
\end{abstract}


\section{Introduction}

Underwater acoustic communications have made impressive progress in the past several decades, ushering us into the age of high-speed acoustic modems and mobile underwater networks. Much of this progress owes to repeated experimental trials that highlighted specific characteristics of acoustic communication channels that distinguish them  from traditional radio channels and  must be taken into account when designing an operational communication system. Advanced signal processing techniques were consequently custom-designed to deal with the extended multipath propagation and Doppler effects, while the design of networking protocols focused on high latency caused by the low speed of sound propagation~\cite{nature,spmag09}. The five orders of magnitude difference in the propagation speed of acoustic waves as compared to electromagnetic waves, together with the wideband nature of acoustic signals, renders traditional radio frequency communication methods inoperable without substantial modifications.

As the field continues to develop, a plethora of new communication techniques are being proposed and documented in the literature. Proof-of-concept, however, remains a point of contention, and only those techniques that have withstood the test of field data  are generally deemed credible by the community. Simulation results obtained using oversimplified channel models meanwhile find their way into the literature, contributing more to misunderstanding than to the advancement of the field. In-water testing thus remains essential, but experimental deployments can be costly, and neither the equipment that they require, nor the signal recordings that they produce, are widely accessible, despite the efforts to enable their effective reuse~\cite{JOE19reuse}. As a result, gathering meaningful test data remains a high barrier to engagement in creative research on underwater acoustic communications, which  remains limited to a few privileged groups.

To bridge this gap, the underwater acoustic communications community must adopt a modeling tool that will capture the important, non-trivial aspects of underwater acoustic propagation, be simple to use but not overly simplistic, and be universally available at no cost. Several such tools have been developed, including those that use impulse responses estimated from field data  to replay a signal through them~\cite{van2017watermark,francois}, and those that focus on statistical channel modeling~\cite{qarabaqi2013statistical,socheleau_parametric_2015}. Nevertheless, none of these tools has emerged as a standard. At this time, there are no standardized channel models, whether statistical or of the replay type, that are widely accepted, yet they are sorely needed by researchers and practitioners around the world, notably those who do not have access to field data.

The need for common acoustic channel models has been repeatedly emphasized in publications, workshops, and meetings~\cite{miljim_mag,editorial,ucomms}. The situation is in stark contrast to terrestrial radio communications, where standard channel models have existed for decades. These models are not only standard in practice but are also \emph{standardized} by international bodies such as the \ac{itu} or the \ac{ieee}. Wide availability of such models has been instrumental in the development of radio communication systems as we know them today, as their adoption enabled widespread research, making it possible for those without extensive experimental facilities to participate in the advancement of the field.

In this paper, we present a library of underwater acoustic channels for purposes of testing and comparing candidate communication, networking, and other signal processing algorithms in a common framework. Our hope is that this library can serve as an initial step towards adoption of a set of standard channel models. While field testing will remain necessary for new advancements, the library will enable the community to meaningfully compare methods within a uniform and \emph{reproducible} framework.

The closest existing effort is the Watermark benchmark~\cite{van2017watermark}, which provides a replay-based simulation framework driven by at-sea impulse response measurements from four sites. Our library expands upon the  Watermark framework in several respects: it covers a substantially wider range of environments and propagation conditions, it includes site-specific noise models that capture both Gaussian and impulsive statistics, and it is distributed as an open-source package with a simple replay interface in three computer languages, \texttt{MATLAB}, \texttt{Python}, and \texttt{Julia}. 

The library is publicly available through an open-access website~\cite{uwachannels_website} and contains a variety of time-varying channel impulse responses and noise characteristics typical of representative environments, including shallow and deep water, short and long range, mobile and fixed platforms, and geographically distinct locations. The channels are modeled using linear time-varying impulse responses extracted from a large volume of field data recordings available to the authors, and include multipath and Doppler effects.  The responses are estimated from field data and stored in matrices showing the evolution of the impulse response over time. Additive noise corresponding to each channel is specified by statistical functions extracted from field data. The library also contains a comprehensive and simple-to-use interface that facilitates application of the channels to user-supplied signals by supporting direct replay. Through this interface, a user can pass an arbitrary  signal through a chosen channel, add noise, and obtain the received signal in a fast and \emph{consistent} manner.

\begin{figure*}[ht]
    \centering
    \begin{tikzpicture}
        \node[anchor=south west, inner sep=0] at (0,0) {\includegraphics[width=\textwidth]{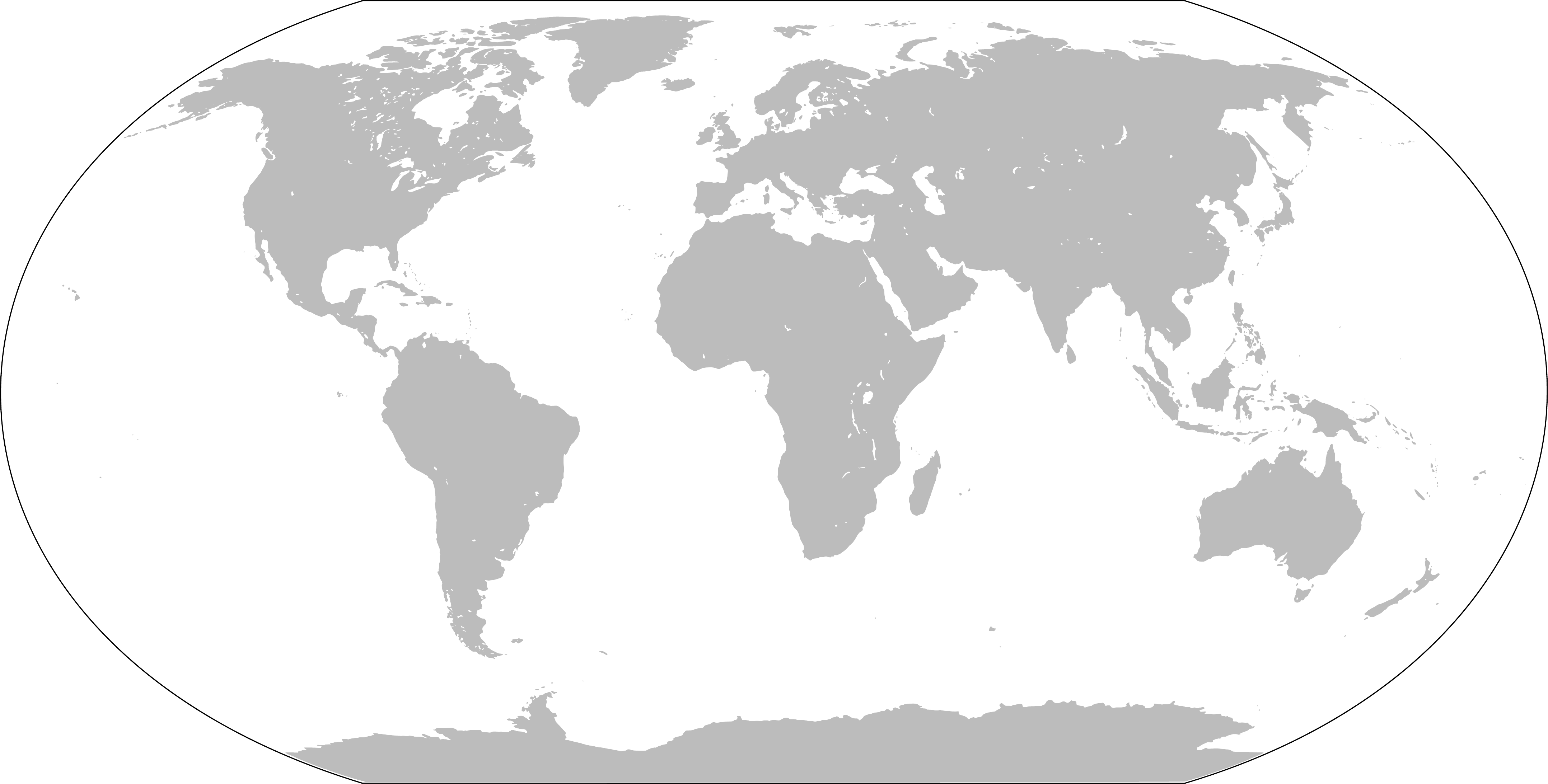}};

        \fill[purple] (5.3, 6.9) ++(-0.15, -0.15) rectangle ++(0.3, 0.3);
        \node[right, font=\small] at (5.6, 6.8) {\purple};

        \draw[blue, line width=2.5pt] (5.4, 7) ++(-0.15, 0) -- ++(0.3, 0);
        \draw[blue, line width=2.5pt] (5.4, 7) ++(0, -0.15) -- ++(0, 0.3);
        \node[right, font=\small] at (5.6, 7.1) {\blue};

        \node[regular polygon, regular polygon sides=3, fill=yellow, draw=yellow, minimum size=14pt, inner sep=0pt] at (0.8, 5.7) {};
        \node[right, font=\small] at (1.0, 5.7) {\yellow};

        \node[regular polygon, regular polygon sides=3, fill=pink, draw=pink, minimum size=14pt, inner sep=0pt, rotate=-90] at (15.2, 6.65) {};
        \node[right, font=\small] at (15.4, 6.65) {\pink};

        \node[regular polygon, regular polygon sides=3, fill=green, draw=green, minimum size=14pt, inner sep=0pt, rotate=180] at (9, 8.2) {};
        \node[above, font=\small] at (9, 8.3) {\green};

        \fill[red] (13.68, 4.75) circle (7pt);
        \node[left, font=\small] at (13.5, 4.75) {\red};

        \node[star, star points=6, star point ratio=2, fill=black, draw=black, minimum size=12pt, inner sep=0pt] at (15.6, 5.35) {};
        \node[right, font=\small] at (15.8, 5.35) {\textcolor{black}{\texttt{black}}};

        \node[regular polygon, regular polygon sides=4, fill=brown, draw=brown, minimum size=12pt, inner sep=0pt, rotate=45] at (2.85, 6.55) {};
        \node[above, font=\small] at (1.85, 6.35) {\brown tx};

        \node[regular polygon, regular polygon sides=4, fill=brown, draw=brown, minimum size=12pt, inner sep=0pt, rotate=45] at (0.8, 5.5) {};
        \node[right, font=\small] at (1.0, 5.45) {\brown rx};

    \end{tikzpicture}

    \caption{Colored markers on this map show the locations of experiments from which the library's channel models were extracted. The \blue and \purple channels, which are close to one another, are in the northern U.S. Atlantic, the \green channel is in Norway, the \red channel in Singapore, the \yellow channel near \hawaii, the \pink channel in Japan, and the \textcolor{black}{\texttt{black}} channel near the Mariana Trench. The \brown channel spans a long range, hence its transmitter (\texttt{tx}) and receiver (\texttt{rx}) locations are marked separately on the map.}
    \label{fig:experiments}
\end{figure*}

Locations where the experimental channels were recorded are shown in Figure~\ref{fig:experiments}. This figure shows the map of the world's oceans, with colored markers placed at the locations of experiments. In the library, and in this paper, we refer to the channels by the color corresponding to the markers on the map.

After landing on the welcome page of the channel repository website~\cite{uwachannels_website}, one can go to the channels, noise, user's guide, or directly to the \texttt{GitHub} repositories, where all of the code resides. The code is available in \texttt{MATLAB}, \texttt{Python}, and \texttt{Julia}. The \texttt{Julia} version is hosted under the \texttt{UnderwaterAcoustics.jl} package~\cite{UnderwaterAcoustics_jl}.

In this paper, we provide the technical details of building the library, along with illustrative examples that explain how to use it. The paper is organized as follows. Section~\ref{sec:overview} provides the list of channels and their basic features, such as location, bandwidth, mobility, and use of array recording. In Section~\ref{sec:algo}, we describe the process of impulse response estimation from field data. Methods used for compressing the channel information for efficient storage within the library are also described here. Section~\ref{sec:noise} contains the description of noise models accompanying each of the channels. In Section~\ref{sec:verif}, we outline the methods used for verifying the accuracy of channel extraction and compression, and discuss  the performance of a communication system using the channel models. Section~\ref{sec:use} is devoted to the details of using the library, including instructions on how to view the channel, how to apply a channel to an arbitrary signal and how to add noise at a desired level. Finally, Section~\ref{sec:gal} contains a gallery of channels, where each channel's details are summarized in a table, and a set of images is provided to illustrate the time-varying impulse responses. We conclude in Section~\ref{sec:concl} with a summary, a call for participation through adding new channels to the library, and an outline of future efforts to enlarge the library with statistical channel models.

\section{Overview of the Channels}
\label{sec:overview}

The channel impulse responses and noise statistics in the library were derived from field data recorded at a variety of locations shown in Figure~\ref{fig:experiments}. The recordings were made during experimental deployments organized and supported by various research groups. While each channel is different, the channel probing signals used in the experiments were of the same type. These signals are derived from \ac{pn} sequences, periodically repeated and modulated onto a single carrier. It is important to note that since the channel impulse responses are estimated from the transmitted signal, the resulting models are valid only within the frequency band of that signal. In other words, when generating a signal to be passed through a certain channel, the signal should conform to the stated frequency band limitation. The noise statistics are similarly estimated in-band. In Table~\ref{tab:params}, we list the parameters of each experiment.

\begin{table*}[h]\centering
\caption{Parameters of the experiments collected in the underwater acoustic channel repository. The symbols $d_T$, $d_R$, and $d_w$ denote the transmitter depth, receiver depth, and water depth, respectively. The parameter $d$ is the transmitter-receiver distance, $f_c$ is the center frequency, $R$ is the symbol rate of the transmitted probe signal, $M$ is the number of array elements, and $\ell$ is the inter-element spacing. An ``x'' indicates that the value varies across different scenarios. For the \green channel, $M$ denotes the number of time diversity channels formed from repeated transmissions on a single hydrophone, with the inter-transmission interval of 10 minutes. }
\begin{tabular}{lllllllllll}
\toprule
Codename              & Location                        & Date                       & Mobility                & $d_{T}$/$d_{R}$/$d_{w}$ {[}m{]} & $d$ {[}km{]}          & $f_c$ {[}kHz{]}       & $R$ {[}ksym/s{]}                & Array                        & $M$                          & $\ell$ {[}m{]}              \\
\midrule
\blue                    & North Atlantic                  & Jun. 2010                  & Mobile            & 30-60/50/100                    & 3-7                   & 13                    & $10^4/2048$                  & Vertical                     & 12                           & 0.12                           \\ \midrule
\red                     & Singapore                       & Nov. 2024                  & Drifting                & 6/4.6/8-20                           & 0.1-0.4                   & 25                    & 9.6                          & Vertical                     & 3                            & 0.8                           \\ \midrule
\multirow{4}{*}{\yellow} & \multirow{4}{*}{Hawaii}         & \multirow{4}{*}{Jul. 2011} & \multirow{4}{*}{Moored} & \multirow{2}{*}{50/50/100}      & 3                     & \multirow{4}{*}{13}   & \multirow{4}{*}{6.25}        & \multirow{4}{*}{Vertical}    & 24                           & 0.05                            \\
\cmidrule{6-6}\cmidrule{10-11}
                        &                                 &                            &                         &                                 & 7                     &                       &                              &                              & 24                           & 0.2                           \\
\cmidrule{5-5}\cmidrule{6-6}\cmidrule{10-11}
                        &                                 &                            &                         & \multirow{2}{*}{50/8.6-65/100}  & 3                     &                       &                              &                              & 16                           & 3.75                         \\
\cmidrule{6-6}\cmidrule{10-11}
                        &                                 &                            &                         &                                 & 7                     &                       &                              &                              & 16                           & 3.75                         \\ \midrule
\multirow{3}{*}{\purple} & \multirow{3}{*}{North Atlantic} & \multirow{3}{*}{Oct. 2008} & \multirow{3}{*}{Moored} & \multirow{3}{*}{11/10/15}       & 0.06                  & \multirow{3}{*}{12.5} & \multirow{3}{*}{$10^4/1536$} & Cross                        & 32                           & 0.0375                         \\
\cmidrule{6-6}\cmidrule{9-11}
                        &                                 &                            &                         &                                 & 0.2                   &                       &                              & Vertical                     & 24                           & 0.05                            \\
\cmidrule{6-6}\cmidrule{9-11}
                        &                                 &                            &                         &                                 & 1                     &                       &                              & Vertical                     & 12                           & 0.12                           \\ \midrule
\green  & Norway & Nov. 2024 & Moored & 20/43/60 & 0.27 & 6 & 4.5 & Time & 64 & - \\ \midrule
\black & Mariana Trench                  & Oct. 2024                  & Moored                  & 8718/6/8720                     & 8.72                  & 18                    & 12.5                         & Planar                     & 8                            & $\geq$ 0.088                            \\ \midrule
\pink                    & Japan                           & Jul. 2022                  & Moored                  & 176/146/x                       & 14                    & 6                     & 4                            & Vertical                     & 24                           & 0.9-1.8                            \\ \midrule
\brown                   & Pacific                          & Nov. 1994                  & Mobile                  & 652/900-1600/x                       & 3250                  & 0.075                 & 0.0375                       & Vertical                     & 20                           & 35                         \\
\bottomrule
\end{tabular}
\label{tab:params}
\end{table*}

\subsection{\blue Channel}

The field data of the \blue channel were recorded during the June \ac{mace} that took place near the coast of Rhode Island in the USA. The  goal of the \ac{mace} experiment was to collect data for research on mobile acoustic communications in water depths of approximately 100~m.
The experimental setup included four transmit elements and a moored 12-element receiver array with an element spacing of 12~cm at a depth of 50~m. The depth of the transmitter varied between 30~m and 60~m.

The acoustic band used in this experiment was 10.5~kHz -- 15.5~kHz. The transmitter was deployed from a moving vessel, while the receiver was anchored. The transmitter moved in a racetrack pattern at a varying speed up to 1.5~m/s, sometimes going towards and sometimes away from the receiver. The transmission distance varied between 3~km and 7~km, and the water depth was about 100~m. The \ac{pn} sequence used for probing the \blue channel had period 2047 and was transmitted repeatedly for approximately one minute, using \ac{bpsk} modulation. The symbol rate  was approximately 5~ksym/s. The center frequency was 13~kHz.

\subsection{\red Channel}

The field data of the \red channel were recorded during the Singapore'24 experiment, and consisted of repetitions of a probe signal transmitted from a drifting boat. The experiment took place near Singapore in November 2024. The signal was captured using a three-channel vertical array with 0.8 m spacing, with the top hydrophone positioned at  4.6~m depth and the transmitter deployed at around 6~m depth. During the drifts, the range between the transmitter and receiver varied between 100~m and 400~m. The bathymetry in the area was not uniform, with water depths ranging from approximately 8~m to 20~m. Additionally, the experiment was conducted near a heavy shipping channel, introducing potential interference and bubbles. The probe signal consisted of 60 repetitions of a \ac{bpsk}-modulated generalized m-sequence of length 8191, transmitted in a frequency band between 20~kHz and 30~kHz and a carrier frequency of 25~kHz. The symbol rate was 9.6~ksym/s.

\subsection{\yellow Channel}

The \yellow channel was extracted from the \ac{kam} experiment, which was conducted off the western side of Kauai, \hawaii, in July 2011. The primary objective of the \ac{kam} experiment was to collect acoustic and environmental data to study the interactions between oceanographic variability, acoustic propagation, and underwater communications. Specifically, the experiment aimed to examine how fluctuations in the ocean environment and source and receiver motion influence variations in the acoustic impulse response between transmitters and receivers, ultimately affecting the design and performance of shallow-water acoustic communication systems. The study focused on short-term fluctuations over time scales of a few to several tens of seconds, which significantly impacted signal reception and degraded communication reliability. By  measuring the environmental conditions during transmission of channel-probing waveforms and communication signals, the experiment sought to assess the feasibility of realistically simulating spatiotemporal channel impulse responses and received communication waveforms. Additionally, it aimed to evaluate the predictability of acoustic communication system performance under different source-receiver configurations and environmental conditions, as well as the short-term predictability of system performance based on real-time signal observations at the receiver.

The signal used to extract the channel impulse responses was a \ac{bpsk}-modulated 4095-bit \ac{pn} sequence with a center frequency of 13~kHz and a symbol rate of 6.25~ksym/s. The  sequence of length 4095 was repeated multiple times. Four multi-channel (array) receive systems were employed in the experiment. The first and second systems used vertical linear arrays with 24 elements, equally-spaced by 5 cm and 20 cm, and transmitter-to-receiver distances of 3~km and 7~km, respectively. The transmitter and receivers were deployed at mid-depth in a 100 m water column. The third and fourth systems used vertical linear arrays with 16 elements, equally-spaced by 3.75~m, and transmitter-to-receiver distances of 3~km and 7~km, respectively. The receivers spanned depths from 8.6~m to about 65~m below the surface in a 100~m water column.

\subsection{\purple Channel}

The data for estimating the \purple channel impulse response and noise statistics were collected during the \ac{space} experiment. \ac{space} was conducted in the fall of 2008 at the Woods Hole Oceanographic Institution's Martha's Vineyard Coastal Observatory, off the south  coast of Martha's Vineyard, Massachusetts. The water depth was 15 m, and the  receiving equipment consisted of five  systems, each with an acoustic receiving array in one of three configurations.  The first configuration consisted of a cross-array with each leg  of the array (horizontal or vertical) having 16 elements with an inter-element spacing of 3.75 cm.  The cross planar arrays were mounted on systems 1 and 2. The other two configurations were vertical line arrays. One configuration had 24 elements with an inter-element spacing of 5 cm (systems 3 and 4) and the other was a 12-element array with an inter-element spacing of 12 cm (system 5). In all arrays, the top element was positioned approximately 3.3 m above the seabed. All systems received the signals transmitted by a single system with a transducer positioned 4 m above the seabed.

The five receive systems were positioned as follows. Systems 1 and 2 were 60 m from the transmitter in the southeast and southwest directions, respectively. Systems 3 and 4 were 200 m from the transmitter in the southeast and southwest directions, respectively. System 5 was 1~km from the transmitter in the southeast direction. The cross arrays on systems 1 and 2 were oriented so that the broadside to the plane of the array was pointed towards the transmitter.

 The acoustic band used in this experiment was approximately 9.2~kHz -- 15.8~kHz. The 4095-bit \ac{pn} sequence was  \ac{bpsk}-modulated, and sent repeatedly. The center frequency was 12.5~kHz, and the symbol rate was approximately 6.51~ksym/s.

\subsection{\green Channel}

The \green channel was recorded during the Norway'24 experiment, which was conducted in an archipelago near the west coast of Norway in November 2024. The main objective of the experiment was to collect hydrophone recordings that could be released into the public domain~\cite{smartocean}. The  channel-probing waveform  was a 60 s long BPSK modulated \ac{pn} sequence with a center frequency of 6~kHz and  a symbol rate of 4.5~ksym/s. This signal  was transmitted over a range of 270~m and received  on a single hydrophone. Unlike with the other channels, where the signals were recorded  over an array of hydrophones, in this experiment the probing waveform was transmitted and recorded every 10 minutes in an otherwise stationary configuration. These repetitions give access to time diversity, with opportunities for processing gain analogous to array processing as used with spatial diversity channels. Time diversity processing is also relevant to practical scenarios in which  a network node stores  a packet it failed to decode, and subsequently combines it with retransmissions. The repository provides access to 64 time diversity slots for the \green channel.

\subsection{\black Channel}

The data for the \black channel were collected during an acoustic communications experiment that took place over a vertical link in the Mariana Trench in October 2024 by the \ac{jamstec}. The transmitter was located  8718~m below the surface, and an eight-element planar array was mounted below the  ship, at a depth of 6~m. The array elements were arranged as vertices of two squares, one inside another, with vertices pointing in the same direction.  The outer  square's side was 264~mm, while the inner square's side was 88~mm.   The signal had a symbol rate of 12.5~ksym/s, and a center frequency of 18~kHz. An 8192-bit binary \ac{pn} sequence was \ac{qpsk} modulated.

\subsection{\pink Channel}

The field data for the \pink channel were collected off of the coast of Kochi Prefecture, Japan, by the \ac{jamstec} in July 2022. The transmission  distance was around 14~km. The  bathymetry ranged from flat to down-slope and up-slope (hence the water depth entry x in the Table~\ref{tab:params}). The nominal water depth was around 200~m. The transmitter was  mounted at about 176~m depth. The receiver array was equipped with 24 elements, grouped into 6 groups of 4 hydrophones.  The groups were spaced by 1.8~m, and the elements within each group were spaced by 0.9~m.  Out of the 24 elements, 21 provided useful signals for channel estimation in one of the three bathymetry configurations. The receiver array was positioned at a depth spanning 146.5~m to 172.5~m. The signal was a  \ac{qpsk}-modulated 8192-bit binary \ac{pn} sequence sent periodically, and the symbol rate was 4~ksym/s. The center frequency was 6~kHz. 

\subsection{\brown Channel}

The field data for the \brown channel were collected in the eastern North Pacific Ocean as part of the \ac{atoc} project in November 1994. The 75-Hz acoustic source was suspended at 652~m depth moored off San Diego, California, in approximately 4000~m deep water. The receiver was a 20-element autonomous vertical line array moored east of Hawaii in about 5300~m deep water, at approximately 3250~km range from the source across the Pacific Ocean. The array elements were spaced 35~m apart, spanning depths from 900~m to 1600~m.  A \ac{bpsk}-modulated 1023-bit \ac{pn} sequence was sent periodically.   The center frequency was 75~Hz, and the symbol rate was 37.5~sym/s.

\section{Channel Extraction and Compression}
\label{sec:algo}

We begin this section by describing the basic method of estimating the channel impulse response from the received signal data. The basic method includes adaptive channel estimation and phase tracking. After outlining the basic method, we move on to discuss the effect of motion-induced delay drift, and the resulting need for delay tracking. We then present a method for delay tracking, or adaptive signal resampling, and incorporate it into the overall time-varying channel impulse response estimation. The overall channel extraction thus includes adaptive channel estimation, phase tracking and delay tracking. The phase and delay tracking are accomplished by the use of a \ac{pll} and a \ac{dll}, respectively.

\subsection{Basic Channel Estimation and Phase Tracking}

Basic channel impulse response estimation from transmitted probes over the bandwidth of interest is based on modeling the equivalent complex baseband received signal as
\begin{equation}
	v(t)=\sum_n d(n)h(t-nT)e^{j\theta(t)}+w(t)
\end{equation}
where $d(n)$ are the transmitted data symbols taken from a \ac{bpsk} or \ac{qpsk} alphabet, $T$ is the symbol interval, $h(t)$ is the complex-valued channel impulse response from which the phase $\theta(t)$ is decoupled, and $w(t)$ is the complex baseband noise. While the term ``channel impulse response'' is sometimes meant to include the phase, we treat the two separately because the phase typically varies much more rapidly than the rest of the channel response, and separate treatment leads to effective estimation of the individual terms. In general, the channel response is also varying with time, but for the moment we will omit explicit labeling for simplicity. However, we will develop an adaptive algorithm that tracks the time-variation of both the channel impulse  response and the phase, and the labeling will then become apparent.

The signal $v(t)$ is obtained by demodulating the raw received signal by the center (carrier) frequency, filtered to band-limit the noise, and sampled at the sampling frequency $f_s=N_s/T=1/T_s$, where $N_s$ is an integer number of samples per symbol and $T_s$ is the sampling period. For example, if raised-cosine pulse shaping is used such that the signal is band-limited to $B=(1+\alpha)/T$, where $\alpha<1$ is the roll-off factor, $N_s=2$ will suffice. Prior to channel impulse response estimation, the signal is synchronized in time such that the cross-correlation between $v(nT)$ and $d(n)$ peaks at lag zero.

Since $v(t)$ is band-limited to $\pm B/2$, the equivalent channel impulse response, as seen within the bandwidth $B$, can be represented by the samples $h(kT_s)$ taken at the sampling rate $f_s=1/T_s>B$. Sampling the received signal at the rate $f_s$ thus yields
\begin{equation}
	v(nT_s)= \sum_k h(kT_s)x(nT_s-kT_s) e^{j\theta(nT_s)} +w(nT_s)
	\label{eq:vsmp}
\end{equation}
where $x(nT_s)$ is the data sequence zero-padded with $N_s-1$ zeros between adjacent data symbols. If $T_s=T/2$, this sequence will include alternating non-zero values, i.e., $d(0), 0, d(1), 0, d(2),0$, etc.

Let us now define the channel impulse response vector as ${\bf h}= \begin{bmatrix}\cdots & h(-T_s) & h(0) & h(T_s) & h(2T_s) & \cdots \end{bmatrix}^\top$. This vector is long enough to capture the salient multipath components, or delay spread. In other words, it  includes all the samples of the channel impulse response that contain appreciable energy, and does not need to be any longer. At this point, it is also convenient to introduce explicit labeling that reflects the time-dependence of the channel response. Specifically, we define the  vector ${\bf h}[n]$ as the vector ${\bf h}$ observed at time $nT_s$, i.e., ${\bf h}[n]= \begin{bmatrix} \cdots & h(-T_s,nT_s) & h(0,nT_s) & h(T_s,nT_s) & \cdots \end{bmatrix}^\top$, where $h(kT_s,nT_s)$ denotes the value of the channel impulse response at delay $kT_s$ and time $nT_s$. The corresponding data vector is defined as ${\bf x}[n] = \begin{bmatrix} \cdots & x(nT_s+T_s) & x(nT_s) & x(nT_s-T_s) & \cdots \end{bmatrix}^\top$. With these definitions, the received signal \eqref{eq:vsmp} can be expressed as
\begin{equation}
	v(nT_s)=\underline v(nT_s)e^{j\theta(nT_s)}+w(nT_s)
\end{equation}
where
\begin{equation}
   \underline v(nT_s)= {\bf h}^\top[n]{\bf x}[n]
\end{equation}
At each sample $n$ (a time increment of $T_s$), as new signal samples are received, the elements of ${\bf x}[n]$ are shifted in position by one element, and one new element is placed on top. If $T_s=T/2$, the new element is either a new data symbol or a 0 (even/odd samples). Meanwhile, the channel response is not expected to change much over a single sampling interval. This allows for efficient channel tracking using conventional adaptive estimation algorithms.

Channel impulse response and phase estimation draw on the idea of estimating the signal $v(nT_s)$ using the underlying estimates $\hat{\mathbf{h}}[n]$ and $\hat \theta(nT_s)$. The estimated signal is formed as
\begin{equation}
	\hat v(nT_s)= \hat{\underline v}(nT_s) e^{j\hat \theta(nT_s)}
	\end{equation}
where
\begin{equation}
 \hat {\underline v} (nT_s)=\hat{\mathbf{h}}^\top [n]{\bf x}[n]
 \label{eq:vest}
\end{equation}
The corresponding estimation error is calculated as
\begin{equation}
	e(nT_s)=v(nT_s)e^{-j\hat \theta(nT_s)}-\hat{\underline v}(nT_s)
\end{equation}

The estimate \eqref{eq:vest} is expressed as an inner product between the conjugate channel vector estimate $\hat {\bf h}^*[n]$ and the vector ${\bf x}[n]$.  Channel estimation can be performed adaptively, using the \ac{mmse} or other estimation criteria, producing a solution of the form

\begin{equation}
	\hat {\bf h}^*[n+1]=\hat {\bf h}^*[n] +{\cal A} \left[{\bf x}[n], e(nT_s) \right]
\end{equation}
where ${\cal A}[{\bf x}, e]$ represents the update from one of a variety of adaptive estimation algorithms operating on the input ${\bf x}$ and the error $e$. For example, if the \ac{lms} algorithm was used, we simply have that ${\cal A}[{\bf x},e]= \mu {\bf x}e^*$, where $\mu$ is the step size. For the channel impulse response estimates  provided in the library, we use either the \ac{lms} algorithm, or the \ac{rls} algorithm.  Note that a new pair of channel impulse response and phase estimates is obtained every $T_s$ seconds.

The estimate $\hat{\bf h}[n]$ obtained in this manner is a fully populated vector. If a sparse channel estimate is desired, thresholding estimates by magnitude or other sparse-estimation methods could be used. Alternatively, a dedicated sparse estimation algorithm, such as matching pursuit, can be used, which can similarly terminate when the residual falls below a pre-specified threshold~\cite{weichang}. The channels provided in the library are the result of full estimation, where no additional sparsing is imposed. In this manner all the information available in the experimental recordings is preserved, while a user can perform additional sparsing as desired.

The phase is updated using a \ac{pll}, whose operation is based on the gradient\footnote{$\Re\{\cdot\}$ and $\Im\{\cdot\}$ denote the real and imaginary parts of a complex number, respectively.}
\begin{equation}
	\psi(nT_s)=-\frac{1}{2} \frac{\partial |e^2(nT_s)|}{\partial \hat \theta}=-{\Im \left[v(nT_s)e^{-j\hat \theta(nT_s)}e^*(nT_s)\right]}
\end{equation}
This gradient, or the instantaneous loop error, is used to update the phase according to
\begin{equation}
	\hat \theta(nT_s+T_s)=\hat \theta (nT_s) +\mathcal{P}\left[\psi(nT_s)\right]
\end{equation}
where $\mathcal{P}[\cdot]$ describes the loop filtering. A second-order loop is specified by
\begin{equation}
	\mathcal{P}[\psi(nT_s)]= K_{f_1}\psi(nT_s)+ K_{f_2}\sum_{i\leq n}\alpha^{n-i} \psi(iT_s)
	\label{eq:pll}
\end{equation}
where $K_{f_1}, K_{f_2}$  and $\alpha$ are the loop parameters. These parameters are typically chosen as $K_{f_2}=K_{f_1}/10$ and $\alpha=1$. The choice of $K_{f_1}$ depends on the power of the signal and the noise, among other factors, and can vary. The values of the  \ac{pll} parameters, as well as those of the  channel estimation algorithm parameters, are stored along with the phase and channel estimates in the corresponding  \texttt{MAT-files}.

\begin{figure}[ht]
    \centering
    \begin{tikzpicture}[
    >=Stealth,
    block/.style={draw, minimum width=2cm, minimum height=1cm, font=\large},
    lbl/.style={font=\normalsize},
]
 
\node[block] (est1) {estimator};
 
\node[lbl, left=1cm of est1, yshift=0.25cm]  (x1) {$x(nT_s)$};
\node[lbl, left=1cm of est1, yshift=-0.25cm] (v1) {$v(nT_s)$};
 
\node[lbl, right=1cm of est1, yshift=0.25cm]  (h1) {$\hat{\mathbf{h}}[n]$};
\node[lbl, right=1cm of est1, yshift=-0.25cm] (p1) {$\hat \theta(nT_s)$};
 
\draw[->] (x1.east) -- (x1.east -| est1.west);
\draw[->] (v1.east) -- (v1.east -| est1.west);
 
\draw[->] (h1.west -| est1.east) -- (h1.west);
\draw[->] (p1.west -| est1.east) -- (p1.west);

\end{tikzpicture}
    \caption{Basic channel extraction takes as the input the  zero-padded transmitted data sequence $x(nT_s)$  and the received signal $v(nT_s)$, and   produces the  time-varying channel estimate $\hat{\mathbf{h}} [n]$ and the phase estimate $\hat{\theta}(nT_s)$.  The operations performed by the estimator block are specified by Eqs.~\eqref{eq:vest}-\eqref{eq:pll}.}
    \label{fig:basic}
\end{figure}
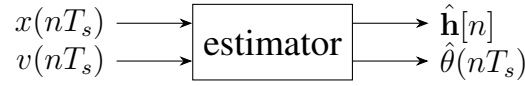

Basic channel extraction is captured  in Figure~\ref{fig:basic}, where the block ``estimator"  performs the operations listed in Eqs.~\eqref{eq:vest}-\eqref{eq:pll}.
For array receivers, a separate estimator is applied for each receiving element. With $M$ elements, each element supplies a different received signal, $v_i(nT_s), i=0,\ldots, M-1$, to its estimator.  Front-end timing synchronization in this case is performed on the first element  and applied across the array. The multiple received signals are thus sampled at the same starting time so as to preserve  the relative time offset  between multiple channel impulse response estimates.

\subsection{Delay Tracking}

The effect of motion is evident not only in the phase, but also in the channel estimate. Specifically, motion causes the channel impulse response to drift over time. When motion persists in the same direction,   it will 
ultimately cause the channel impulse response to slip outside of the pre-allocated window set by the length of the channel vector. If one were to capture all the multipath energy as time goes on, one would need to enlarge this window, i.e., specify a longer channel impulse response vector. A longer vector will in turn require an adjustment of the estimator parameters (lower step size for LMS, higher forgetting factor for RLS), ultimately reducing the ability to track fine scale channel impulse response fluctuations that are not directly modeled as either phase rotation or time dilation or compression. In addition, a longer channel impulse response vector will also require a wider matrix to store the channel. Meanwhile, the delay spread remains unchanged.
In other words, to accommodate the motion-induced drift, the matrix used for storing the channel responses would need to have the delay dimension greater than the minimum dictated by the multipath spread. This problem can be eliminated by compensating for the time-varying delay in the received signal prior to channel estimation.

To address the problem of delay drift, and the resulting storage issues in long-term channel estimation, we incorporate an adaptive resampling mechanism. Adaptive resampling accounts for delay tracking and automatic delay adjustment of the signal at the input to the channel estimator. In order to describe this method, we first take a step back to take a detailed look at the structure of the channel.

Focusing specifically on the path drift in delay, and ignoring the intra-path spreading around the nominal delay for simplicity, the received passband signal is modeled as
\begin{equation}
	r(t)=\sum_p h_p(t) s(t-\tau_p(t))+n(t)
\end{equation}
where $s(t)=\Re\{u(t)e^{j2\pi f_ct}\}$ is the transmitted passband signal,  $u(t)=\sum_k d(k)g(t-kT)$ is the transmitted baseband signal with the transmitter pulse shape $g(t)$, and $n(t)$ is the additive noise. The channel is modeled as having multiple propagation paths, with $h_p(t)$ and $\tau_p(t)$ representing, respectively, the gain and delay observed on the $p$-th path at time $t$. Expressing the received signal as $r(t)=\Re\{v(t)e^{j2\pi f_c t}\}$, we obtain the equivalent baseband received signal
\begin{equation}
	v(t)=\sum_p h_p(t)u(t-\tau_p(t))e^{-j 2\pi f_c \tau_p(t)} +w(t)
\end{equation}
where $w(t)$ is the complex baseband noise.

The time-varying delays are modeled as
\begin{equation}
	\tau_p(t)=\tau_p+\Delta \tau_p(t)
	\label{eq:taup1}
\end{equation}
where $\tau_p$ is the initial delay and $\Delta \tau_p(t)$ is a time-varying component. The time-varying component is further split into two terms,
\begin{equation}
	\Delta \tau_p(t)=\Delta \tau(t)+\epsilon_p(t)
	\label{eq:taup2}
\end{equation}
where $\Delta \tau(t)$ represents a term common to all the paths, and $\epsilon_p(t)$ represents the path-specific deviation. When the transmitter and receiver are moving towards or away from each other such that all the paths are shortening or lengthening, $\Delta \tau (t)$ captures the dominant delay drift, while the path-specific terms $\epsilon_p(t)$ capture the small remaining deviations. In such cases, this type of modeling enables separation of the dominant phase distortion from the rest of the channel response, which, as we shall see, is advantageous for channel tracking.

We now define the time-varying channel phase as
\begin{equation}
	\theta(t)=-2\pi f_c\Delta \tau (t)
\end{equation}
and the time-varying baseband channel response as
\begin{equation}
	h(\tau,t)= \sum_p \tilde h_p(t)g(\tau-\tau_p(t))
	\label{eq:h}
\end{equation}
where $\tilde h_p(t)=h_p(t) e^{-j2\pi f_c(\tau_p+\epsilon_p(t))} $. As before, $\tau$ and $t$ represent the delay and time, respectively.

With these definitions, the received signal is expressed as
\begin{equation}
	v(t)=\sum_n d(n) h(t-nT,t)e^{j\theta(t)}+w(t)
	\label{eq:vt}
\end{equation}
Note that in mobile channels, the temporal change in $h(\tau,t)$ is typically (much) slower than the change caused by the varying phase $\theta(t)$.

While basic channel estimation relied on the same input-output relationship as seen in Eq.~\eqref{eq:vt}, it assumed no specific structure for either the channel $h(\tau,t)$ or the phase $\theta(t)$. In contrast, we will now exploit the link between these two quantities that is provided by  the time-varying delays.

Specifically, if we define the drift-free channel response as
\begin{equation}
	\underline{h}(\tau,t)=\sum_p \tilde h_p(t) g(\tau-(\tau_p+\epsilon_p(t)))=h(\tau+\Delta \tau(t),t)
\end{equation}
the received signal can be expressed as
\begin{equation}
	v(t)=\sum_n d(n) \underline{h}(\tau-\Delta \tau(t)-nT,t)e^{-j2\pi f_c\Delta \tau (t)}+w(t)
	\label{eq:veps}
\end{equation}
This input-output relationship points to the basic idea of drift-free channel extraction: If one could compensate for the time-varying delay in the received signal, the channel that would remain to be estimated is the drift-free one, $\underline{h}(\tau,t)$, not the drifting channel $h(\tau,t)$.

To illustrate the process, it is useful to express the drift over some short interval of time as $\Delta \tau(t)=a(t)\cdot t$, where $a(t)$ is the Doppler scaling factor. In practical situations, the Doppler scaling factor changes slowly as compared to the sampling interval $T_s$. Hence, $a(t)$ can be approximated as piece-wise constant, $a(t)=a_n$ for $t\in {\cal T}_n$, where ${\cal T}_n$ is a short interval of time surrounding the time $t_n=nT_s$, i.e., the interval of time beginning right after $t_{n-1}$ and ending at $t_n$,   ${\cal T}_n = (t_{n-1},t_n]$, and $a_n\approx a_{n-1}$. Resampling the signal $v(t)$ to compensate for time dilation or compression in this interval now yields
\begin{equation}
	v\left(\frac{nT_s}{1-a_n}\right) = \sum_k x(kT_s) \underline{h}(nT_s-kT_s,\,nT_s) e^{-j\frac{a_n}{1-a_n}2\pi f_c nT_s} + z(nT_s)
	\label{eq:van}
\end{equation}
where $z(nT_s)$ is noise. Denoting the resulting phase by
\begin{equation}
	\varphi(nT_s)=-2\pi f_c \frac{a_n}{1-a_n}nT_s
\end{equation}
the time-scaled signal \eqref{eq:van} can also be expressed as
\begin{equation}
	v\left(nT_s-\frac{\varphi(nT_s)}{2\pi f_c}\right) = \sum_k x(kT_s) \underline{h}(nT_s-kT_s,\,nT_s) e^{j\varphi(nT_s)} + z(nT_s)
	\label{eq:vph}
\end{equation}

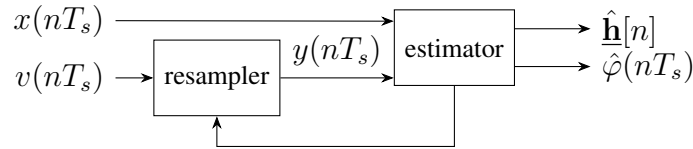
\begin{figure}[h]
    \centering
    \begin{tikzpicture}[
    >=Stealth,
    block/.style={draw, minimum width=1.3cm, minimum height=1cm, font=\large},
    lbl/.style={font=\normalsize},
    xscale=0.75,
]

\node[block, xshift=2cm] (est) {\small estimator};
\node[block, left=1.5cm of est, yshift=-0.4cm] (resamp) {\small resampler};
 
\node[lbl, left=0.5cm of resamp] (v2) {$v(nT_s)$};
 
\node[lbl, anchor=east] at (v2.east |- {$(resamp.north)+(0,0.25cm)$}) (x2) {$x(nT_s)$};
 
\node[lbl, right=1cm of est, yshift=0.25cm]  (h2) {$\underline{\hat{\mathbf{h}}}[n]$};
\node[lbl, right=1cm of est, yshift=-0.25cm] (p2) {$\hat\varphi(nT_s)$};
 
\draw[->] (v2.east) -- (resamp.west);
 
\draw[->] (resamp.east) -- node[above, lbl] {$y(nT_s)$} ($(est.west)+(0,-0.4cm)$);

\draw[->] (x2.east) -- ($(est.west)+(0,0.35cm)$);
 
\draw[->] (h2.west -| est.east) -- (h2.west);
\draw[->] (p2.west -| est.east) -- (p2.west);
 
\draw[->] (est.south) -- ++(0,-0.8cm) -| (resamp.south);
 
\end{tikzpicture}
    \caption{Channel extraction with delay tracking takes as the input the zero-padded transmitted data sequence $x(nT_s)$ and the received signal $v(nT_s)$. The received signal is first resampled, producing $y(nT_s)$, which is then fed along with $x(nT_s)$ to the  estimator, identical as the one of Figure~\ref{fig:basic}. The estimator produces the time-varying, drift-free channel estimate $\hat{\underline{\mathbf{h}}}[n]$ and the phase estimate $\hat{\varphi}(nT_s)$. The phase estimate is fed back to the resampling block to adaptively  correct for the time-varying delay drift.}
    \label{fig:full}
\end{figure}

Imagining for a moment that the phase $\varphi(nT_s)$ is known and that time-scaling (resampling) is performed as above, the signal sample $v\left(nT_s-\frac{\varphi(nT_s)}{2\pi f_c}\right)$ can be used to update the channel estimate $\hat {\bf \underline{h}}[n]$
and the phase estimate $\hat \varphi(nT_s)$. This procedure will yield
$\hat {\bf \underline{h}}[n+1]$
and $\hat \varphi(nT_s+T_s)$. The phase $\hat \varphi(nT_s+T_s)$ can then be used to generate the next signal sample $v\left((n+1)T_s-\frac{\hat \varphi((n+1)T_s)}{2\pi f_c}\right)$, and so on. The process is illustrated in Figure~\ref{fig:full}. The estimator block in this figure is identical  to the one used before in Figure~\ref{fig:basic}, but it takes as its input the resampled received signal and produces as its output the estimate of the drift-free channel ${\bf \underline{h}}[n]$ and the modified phase $\varphi(nT_s)$.

In practice, resampling can be implemented using a simple linear interpolation between two nearest samples of the input signal $v(t)$.
In particular, using the estimate $\hat \varphi(nT_s)$ instead of the true phase $\varphi(nT_s)$, and applying linear interpolation in lieu of full band-limited resampling, we obtain
\begin{equation}
	v\left(nT_s-\frac{ \varphi(nT_s)}{2\pi f_c}\right)\approx{\cal L} \left[ v\left(nT_s-\frac{\hat \varphi(nT_s)}{2 \pi f_c}\right) \right]
\end{equation}
where ${\cal L}[\cdot]$ denotes the linear interpolation (linear resampling) function given by
\begin{equation*}
	{\cal L}[v(t(n))]=(1-\alpha(n))v(t_L(n))+\alpha(n) v(t_R(n))
\end{equation*}
with
\begin{align}
	t_L(n) &= \left\lfloor \frac{t(n)}{T_s} \right\rfloor T_s, \quad
	t_R(n) = \left\lceil \frac{t(n)}{T_s} \right\rceil T_s, \notag \\
	\alpha(n) &= \frac{t(n)-t_L(n)}{T_s}
	\label{eq:rsmp}
\end{align}
where $\lfloor \cdot \rfloor$ and $\lceil \cdot \rceil$ denote the floor and ceiling operators, respectively. 

The signal
\begin{equation}
	y(nT_s)={\cal L}\left[v\left(nT_s-\frac{\hat \varphi(nT_s)}{2 \pi f_c}\right)\right]
	\label{eq:a1}
\end{equation}
that approximates the desired signal $v\left(nT_s-\frac{ \varphi(nT_s)}{2\pi f_c}\right)$ is now used to estimate the channel and the phase as before. The resulting estimates are those of the drift-free channel ${\bf \underline{h}}[n]$ and the modified phase $\varphi(nT_s)$:
\begin{align}
    \hat {\underline y} (nT_s) &=  \underline{\hat{\mathbf{h}}}^\top [n]\,\mathbf{x}[n] \\
	e(nT_s) &= y(nT_s)\,e^{-j\hat{\varphi}(nT_s)} - \hat {\underline y} (nT_s) \\
	\underline{\hat{\mathbf{h}}}^*[n+1] &= \underline{\hat{\mathbf{h}}}^*[n] + \mathcal{A}\!\left[\mathbf{x}[n],\, e(nT_s)\right] \\
	\hat{\varphi}(nT_s+T_s) &= \hat{\varphi}(nT_s) + \mathcal{P}\!\left[-\Im\left\{y(nT_s)\,e^{-j\hat{\varphi}(nT_s)}\,e^*(nT_s)\right\}\right]
	\label{eq:a2}
\end{align}

Expressions \eqref{eq:a1} through \eqref{eq:a2} define the full algorithm that was used to build the library. This algorithm incorporates either \ac{lms}-based or \ac{rls}-based adaptive channel estimation, phase and delay tracking. Note that the phase estimate should be initialized with $\hat \varphi(0)=0$. For array reception, this process is applied independently to each element of the array, with inter-element timing preserved by synchronizing  all the elements to the same starting time (e.g., that of the first element).

\begin{figure*}[ht]
    \centering
    \begin{subfigure}[b]{0.32\textwidth}
        \centering
        \includegraphics[width=\textwidth]{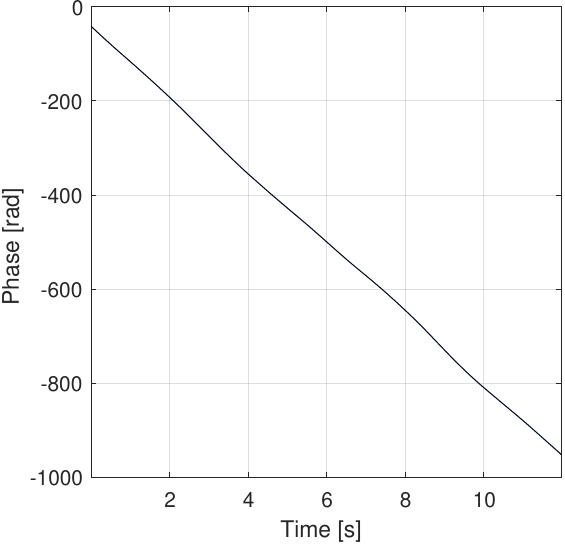}
        \caption{Phase}
        \label{fig:phase}
    \end{subfigure}
    \hfill
    \begin{subfigure}[b]{0.32\textwidth}
        \centering
        \includegraphics[width=\textwidth]{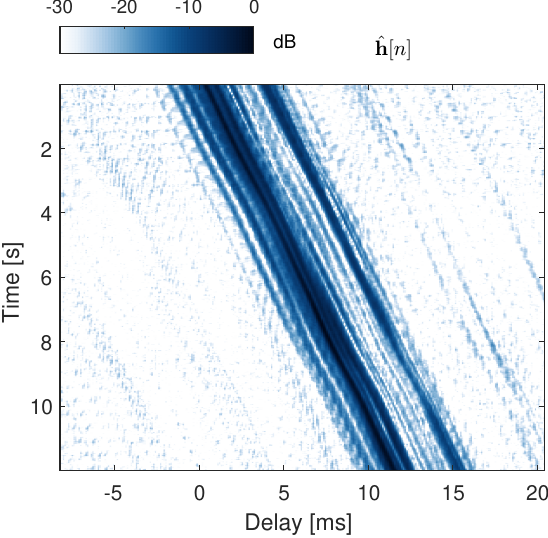}
        \caption{Magnitude (basic)}
        \label{fig:slanted}
    \end{subfigure}
    \hfill
    \begin{subfigure}[b]{0.32\textwidth}
        \centering
        \includegraphics[width=\textwidth]{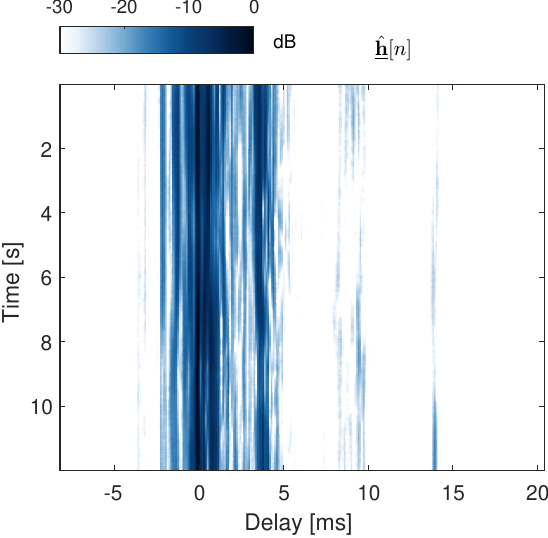}
        \caption{Magnitude (with delay tracking)}
        \label{fig:straightened}
    \end{subfigure}

    \caption{Channel estimation results for a segment of the \blue channel. (a) Phase $\hat{\theta}(nT_s)$ observed on the first element of the receiving array. The considerable variation of phase  over time corresponds to motion at about 1.4~m/s, which agrees with the ship's movement records. (b)~Magnitude of the channel estimate vector $\hat{\mathbf{h}}[n]$ obtained from basic estimation. This  channel estimate drifts over time due to the transmitter motion. (c) Magnitude of the drift-free channel estimate vector $\underline{\hat{\mathbf{h}}}[n]$ obtained with  delay tracking engaged. The drift visible in~(b) is clearly suppressed.}
    \label{fig:channel_extraction}
\end{figure*}

Figure~\ref{fig:channel_extraction} illustrates  the result of  channel extraction applied to one segment of the \blue channel. Shown in the left panel is the phase  $\hat \theta(nT_s)$ whose slope indicates uni-directional motion at a relatively constant speed. In addition,  the figure shows a heat map of the channel matrix obtained without delay compensation (basic method, middle panel), and with delay compensation (right panel). As one can see, without delay compensation, the channel response drifts over time, resulting in the slant of the dark lines across the plot. This slant, or drift, is caused by the same motion that is evident in the phase, and can eventually cause the channel response to slip outside of the allotted delay window (30~ms in this figure).  With delay compensation, the channel response no longer drifts over time. As a result, the channel estimate vector  needs to be only as long as the multipath spread of the channel dictates. In this example,  high-energy multipath is contained within about 7~ms, while all low-energy arrivals are safely captured within about 17~ms. 

The channel estimation algorithms, as well as utility functions that are used to generate plots such as those of Figure~\ref{fig:channel_extraction}, are hosted within the library under the \texttt{GitHub} entry~\cite{uwachannels_website}.

\subsection{Compression}

Channel impulse response information is stored in a matrix whose rows contain the channel vectors $\hat {\bf \underline{h}} [n]$. The number of rows grows with the length of the channel observation time, and can become quite large with data records lasting up to several hours. Compression is thus essential for efficient storage within the library.

Preventing the drift in the channel response slows down the apparent time-variability and allows for efficient compression (reduction of the number of rows that need to be stored). The minimum sampling rate in time that can be used to represent the channel can be determined from the channel's  scattering function, i.e., the power spectral density, or the Doppler power spectrum, evaluated at each delay. The scattering function is defined 
as the Fourier transform of the  auto-correlation  $R_{\underline{h}}(\tau,\Delta t)=\mathbb{E}\{\underline{h}(\tau,t+\Delta t)\underline{h}^*(\tau,t)\}$ taken across the time lag $\Delta t$,  $S_{\underline{h}}(\tau,\nu)={\cal F}[R_{\underline{h}}(\tau,\Delta t))]$. Alternatively, if insufficient data is available to determine the auto-correlation, the  delay-Doppler spreading function can be used to assess  the minimum sampling rate in time needed to represent the channel. The delay-Doppler spreading function is  defined as the Fourier transform of the channel impulse response $\underline h (\tau,t)$  taken across the time $t$, $D_{\underline{h}}(\tau,\nu)={\cal F}[\underline h(\tau,t)]$.  

\begin{figure}[ht]
	\centering\includegraphics[width=0.45\textwidth]{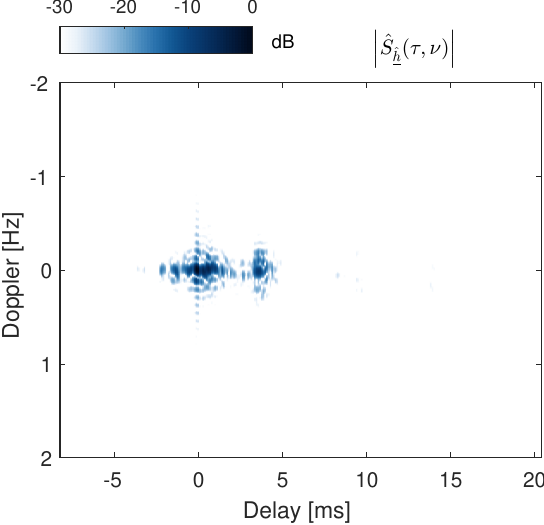}
	\caption{Scattering function of the \blue channel of  Figure~\ref{fig:straightened}.}
	\label{fig:dopp}
\end{figure}

As an example, Figure~\ref{fig:dopp} shows the  estimated scattering function (magnitude) of the \blue channel response $\hat {\bf \underline{h}}[n]$ of Figure~\ref{fig:straightened}.  From this figure, we see that the maximum spectral occupancy is limited to less than $\pm 1$~Hz, implying that a sampling rate of 2 samples/second would suffice for this channel. The same conclusion is reached if one looks at the delay-Doppler function $D_{\underline{\hat h}}(\tau,\nu)$.  Compared to the sampling rate of roughly 10~kHz that was used in~Figure~\ref{fig:straightened},  a compression ratio of 5000 is implied, which is quite substantial.

Compression is applied to all the channels stored in the library. Along with the compressed channel, which is also normalized to provide a unit average squared norm, each library entry contains the corresponding phase $\hat \varphi(nT_s)$, which is stored at the original sampling rate $f_s$, and can be used to reconstruct the full channel. In cases where delay tracking is not applied, the phase $\hat \theta(nT_s)$ is stored.  

Channel reconstruction  involves three steps: decompressing the channel matrix, introducing the time-varying phase, and introducing the time-varying delay drift. The functions that implement channel reconstruction are available for download, and can be applied to any of the channels stored in the library. Details of this process are described in Section~\ref{sec:use}.

\section{Noise}
\label{sec:noise}

While a simple additive white Gaussian noise model is commonly used in simulating underwater communication systems, it fails to capture the true nature of the noise observed in ocean acoustics. Ambient noise in the ocean is typically colored, with a frequency-dependent power spectral density~\cite{urick1984ambient}. A white noise assumption poorly approximates reality for any wideband underwater acoustic system. Underwater ambient noise also has directionality and therefore exhibits spatiotemporal correlation across hydrophones in multichannel systems. Even in the case of isotropic noise, the wideband nature of communication systems can lead to noise correlations across receiving channels. While the Gaussian noise assumption holds in many underwater environments, warm shallow waters exhibit a significant amount of non-Gaussian noise due to snapping shrimp~\cite{chitre2006optimal}. The noise in polar regions is also often non-Gaussian due to ice cracking and bubbles emanating from melting glaciers and icebergs~\cite{dwyer1983technique}. One way to avoid making simplifying assumptions is to use recorded ambient noise from the environment of interest as additive noise samples for simulation. This, however, requires long uncontaminated noise recordings from those environments to avoid repeating noise samples, and is often impractical. 

Instead of providing long ambient noise recordings in the channel library, we adopt an approach to generate synthetic noise samples with a probability distribution and power spectral density similar to that observed in the environment of interest. This allows us to generate either Gaussian or impulsive noise depending on the environment of interest. It also allows us to accurately model the spatiotemporal dependency between samples at different times and on different hydrophones. The exception is the \brown channel, whose recordings  contained only periods with active signal transmissions. Its ambient noise parameters could thus not be estimated, and no site-specific noise model is provided for this channel. Nonetheless,  a generic colored model can be used instead. 

The generation of the noise samples is controlled by several model parameters. These parameters are obtained from noise recordings from the environment and stored in the library. When selecting a channel to use from the library, the corresponding noise parameters may be chosen to generate noise. The noise can then be  scaled and added to the received signal for a specified \ac{snr}.

\subsection{Gaussian noise}

A Gaussian random process is fully characterized by its mean and covariance~\cite{edition2002probability}. Since a hydrophone measures dynamic pressure, the noise is, by definition, zero-mean. The problem of generating synthetic ambient noise then reduces to estimation of the noise covariance and generation of Gaussian random noise with the same covariance.

We consider a system with $M$ receive hydrophones (channels) operating at a sampling rate of $f_s$ samples per second. Let $\tilde{n}_i(t)$, $i=0,\ldots,M-1$, denote the noise recorded on hydrophone $i$ and bandpass-filtered over  the same acoustic band occupied by  the communication signals used to extract the channel impulse responses. We then obtain an estimate of the band-filtered noise covariance by computing sample covariance~\cite{std1968}

\begin{equation}
	\hat{R}_{ij}(\Delta T_s) = \frac{1}{N-2L-1}\sum_{n=L+1}^{N-L} \tilde{n}_{i}(nT_s) \tilde{n}_{j}(nT_s+\Delta T_s) 
    \label{eq:est_noise_cov}
\end{equation}
where the indices $i, j = 0, \cdots,M - 1$, the length $N$ captures  the total number of time samples in the measured noise sequence, $\Delta$ is the discrete time lag, $\Delta = -L,\ldots,L$, and $L$ is the maximum discrete time lag with non-negligible covariance.

Generating noise with the same covariance \eqref{eq:est_noise_cov} involves two steps: (1) generation of standard spatiotemporal white Gaussian noise, and (2) spatiotemporal filtering of the white noise to generate colored Gaussian noise with the desired covariance. The filter coefficients, also referred to as the mixing coefficients  for step (2), can be directly obtained by  decomposing  the covariance matrix, or through an optimization procedure described in~\cite{chitre2024ucomms}. The synthetic ambient noise samples for the $i$-th receiver are generated as

\begin{equation}
	\hat n_{i}(nT_s) = \sum_{j=0}^{M-1}\sum_{k=0}^{L} \beta_{ij}(kT_s)\eta_{j}(nT_s-kT_s)
    \label{eq:mixing}
\end{equation}
where $i = 0, \cdots, M - 1$, $\beta_{ij}(kT_s)$ are the mixing coefficients, and $\eta_{j}(nT_s) \sim \mathcal{N}(0,1)$ are standard independent Gaussian random samples, corresponding to band-limited white Gaussian noise at a bandwidth corresponding to that of the channel.

\subsection{Impulsive noise}

The heavy-tailed \emph{stable} family of distributions has been used extensively in the literature to model impulsive data~\cite{nolan2020stable,nikias1995signal}, and the $\alpha$-stable sub-Gaussian ($\alpha$SG) random process has been shown to model ambient noise in tropical shallow waters well~\cite{mahmood2015modeling}. The stable distribution is a generalization of the Gaussian distribution and shares many desirable features of the Gaussian distribution. When the characteristic exponent $\alpha$ is 2, the stable distribution reduces to the Gaussian distribution. It is therefore not surprising that the method for Gaussian noise can be generalized to impulsive noise~\cite{Chitre2025}.

The $\alpha$SG random process is fully characterized by its characteristic exponent $\alpha$ and the covariance of an underlying multivariate Gaussian distribution. Characteristic exponent $\alpha$ can be estimated using a simple fractile estimator~\cite{chambers1976method}. While one could estimate the underlying Gaussian covariance using moment-based or covariation-based estimators~\cite{kring2009estimation} and then generate $\alpha$SG covariates using the method described in~\cite{mahmood2017generating}, the resulting procedure is computationally expensive. Instead, the method described in~\cite{Chitre2025} uses a procedure similar to that for Gaussian noise using~(\ref{eq:mixing}), but drawing $\eta_{j}(nT_s) \sim \mathcal{S}_\alpha(0,1/\sqrt{2})$ as symmetric $\alpha$-stable random variates with a scale parameter of $1/\sqrt{2}$ to ensure that the $\alpha$-stable distribution reduces to a standard Gaussian when $\alpha=2$. The stability property of stable distributions guarantees that the linear combination of $\alpha$-stable random variates is $\alpha$-stable, leaving us with the problem of determining mixing coefficients $\beta_{ij}(kT_s)$ to ensure that the underlying Gaussian covariance of the generated noise matches that of the ambient noise samples. The optimization procedure in~\cite{chitre2024ucomms} is generalized to $\alpha$SG random process in~\cite{Chitre2025}, and provides a recipe for computing $\beta_{ij}(kT_s)$.

Once  $\alpha$ and $\beta_{ij}(kT_s)$ have been determined from ambient noise recordings, their values are stored  in the channel library. When we require noise samples for simulation, we draw $\eta_{j}(nT_s) \sim \mathcal{S}_\alpha(0,1/\sqrt{2})$ and mix them using~(\ref{eq:mixing}) to generate impulsive noise samples with a distribution similar to that observed in the noise recordings.

In Table~\ref{tab:noise}, we specify the mapping between channel impulse responses and the estimated noise parameters. The \blue, \yellow, \purple, \green, \black, and \pink channels correspond to colored Gaussian noise, whereas the \red channel corresponds to colored impulsive noise. The \brown channel has no site-specific model, but generic colored Gaussian noise can be used here.

\begin{table}[ht]
\centering
\caption{Channel Repository Noises Mapping}
\begin{tabular}{l l l}
\toprule
Codename & Channel(s) & Noise Source(s) \\
\midrule
\blue
    & \texttt{blue\_1 - 20}
    & \texttt{blue\_noise} \\
\midrule
\red
    & \texttt{red\_1 - 4}
    & \texttt{red\_noise} \\
\midrule
\multirow{4}{*}{\yellow}
    & \texttt{yellow\_1 - 3}
    & \texttt{yellow\_noise\_1} \\
    & \texttt{yellow\_4}
    & \texttt{yellow\_noise\_2} \\
    & \texttt{yellow\_5}
    & \texttt{yellow\_noise\_3} \\
    & \texttt{yellow\_6}
    & \texttt{yellow\_noise\_4} \\
\midrule
\multirow{3}{*}{\purple}
    & \texttt{purple\_1 - 5}
    & \texttt{purple\_noise\_1 - 5} \\
    & \texttt{purple\_6 - 10}
    & \texttt{purple\_noise\_1 - 5} \\
    & \texttt{purple\_11 - 15}
    & \texttt{purple\_noise\_1 - 5} \\
\midrule
\green
    & \texttt{green}
    & \texttt{green\_noise} \\
\midrule
\black
    & \texttt{black}
    & \texttt{black\_noise} \\
\midrule
\pink
    & \texttt{pink\_1 - 3}
    & \texttt{pink\_noise\_1 - 3} \\
\midrule
\brown
    & \texttt{brown}
    & - \\
\bottomrule
\end{tabular}
\label{tab:noise}
\end{table}

\subsection{Illustration of Noise Models}

For all the channels, the model parameters were estimated from ambient recordings of varying duration collected during intervals in which no signal transmissions occurred. For the Gaussian noise models, the recordings were bandpass-filtered to the same frequency band as the corresponding communication signals prior to model training. For the impulsive noise model, the recording was bandpass-filtered over a wider frequency band prior to training.

\begin{figure*}[h]
    \centering
    \begin{subfigure}[b]{1\textwidth}
        \centering
        \begin{subfigure}[b]{0.32\columnwidth}
            \includegraphics[width=1\textwidth]{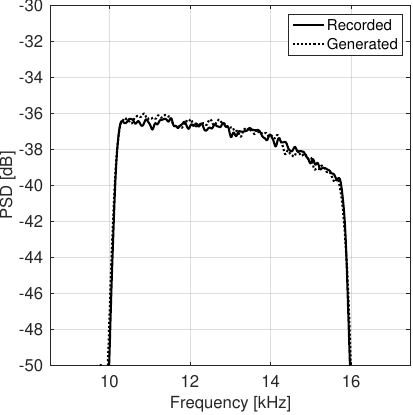}
            \caption{PSD.}
            \label{fig:noise_a}
        \end{subfigure}
        \begin{subfigure}[b]{0.32\columnwidth}
            \includegraphics[width=0.98\textwidth]{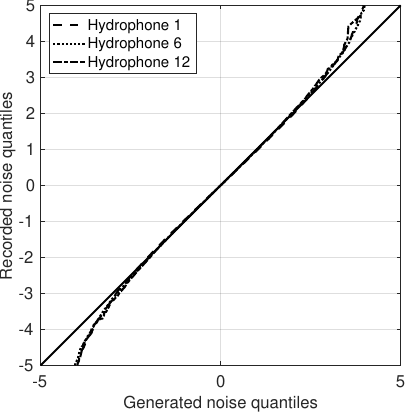}
            \caption{Q-Q plot.}
            \label{fig:noise_b}
        \end{subfigure}
        \begin{subfigure}[b]{0.32\columnwidth}
            \includegraphics[width=1\columnwidth]{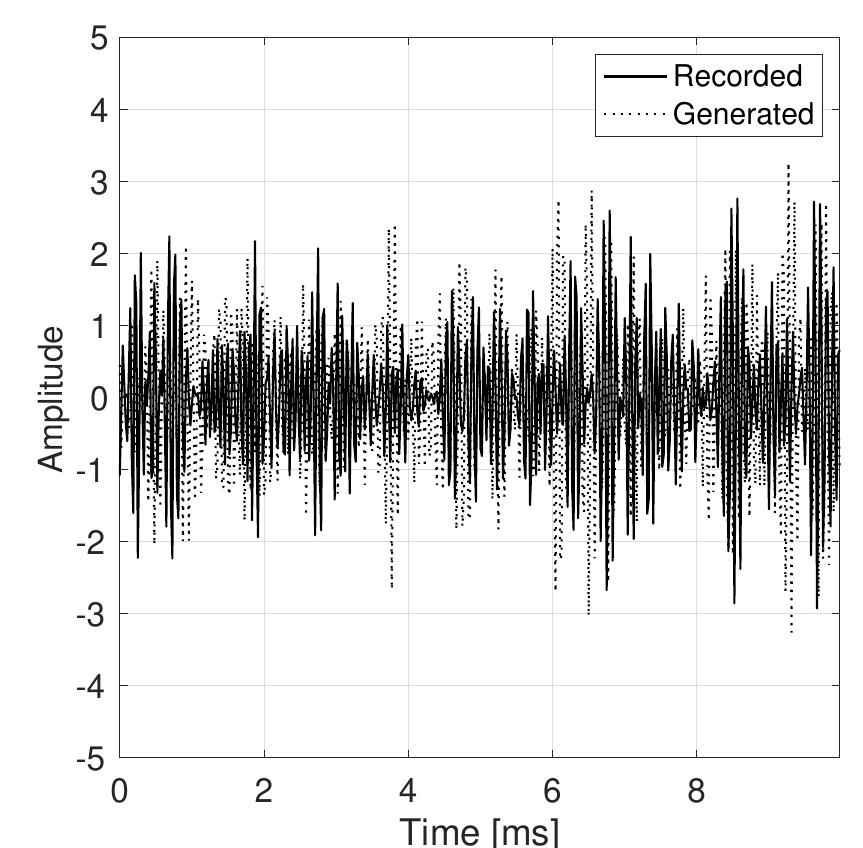}
            \caption{Noise samples.}
            \label{fig:noise_d}
        \end{subfigure}
    \end{subfigure}
    
    \begin{subfigure}[b]{1\textwidth}
        \centering
        \begin{subfigure}[b]{0.32\columnwidth}
            \includegraphics[width=0.98\textwidth]{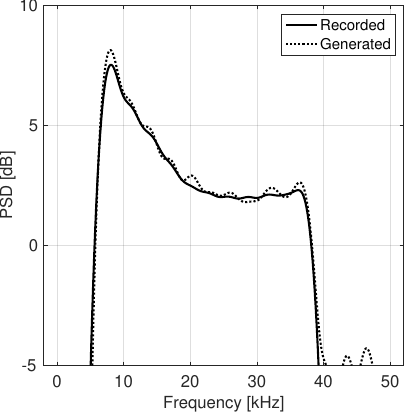}
            \caption{Pseudo-PSD.}
            \label{fig:noise_e}
        \end{subfigure}
        \begin{subfigure}[b]{0.32\columnwidth}
            \includegraphics[width=1\textwidth]{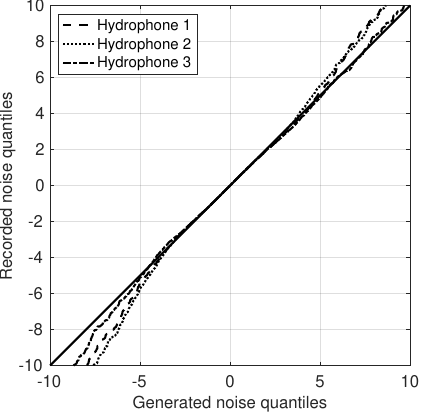}
            \caption{Q-Q plot.}
            \label{fig:noise_f}
        \end{subfigure}
        \begin{subfigure}[b]{0.32\columnwidth}
            \includegraphics[width=1\textwidth]{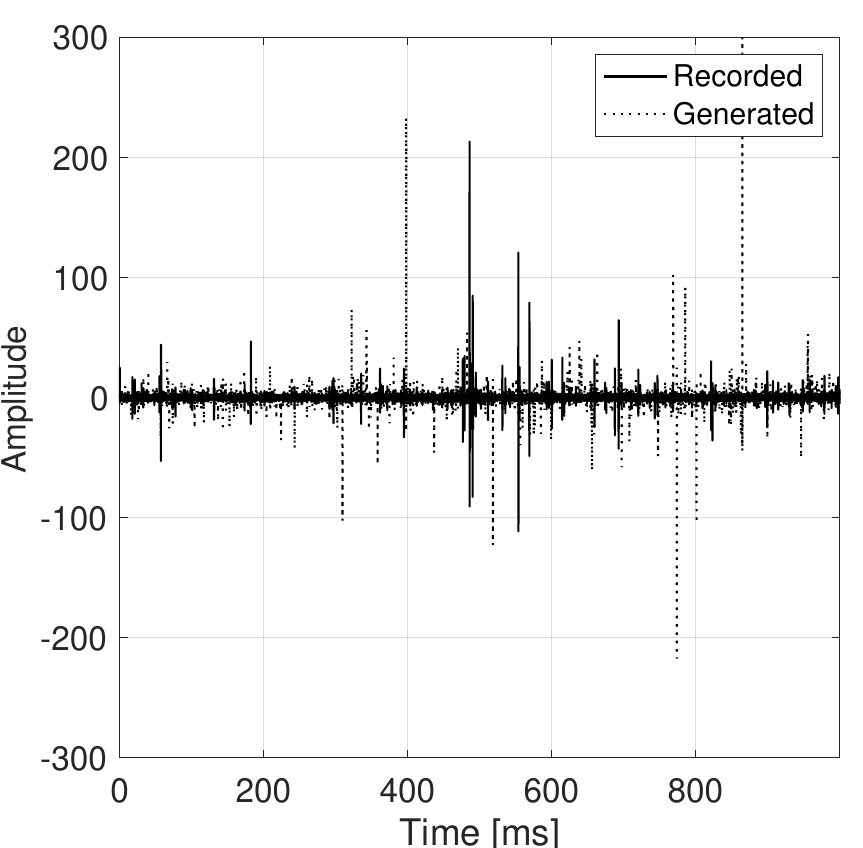}
            \caption{Noise samples.}
            \label{fig:noise_h}
        \end{subfigure}
    \end{subfigure}
    \caption{Ambient noise characterization for the \blue channel (top row) and the \red channel (bottom row). The left column compares the spectral characteristics of the recorded and generated noises, the middle column presents the Q-Q analysis, and the right column compares the recorded and generated time-domain noise samples. The close agreement between the recorded and generated noise confirms the accuracy of the proposed noise model in capturing the statistical and spectral properties of the ambient noise.}
    \label{fig:noise_panel}
\end{figure*}

The noise model parameters \(\beta_{ij}(kT_s)\) were estimated using $L=64$ for the \red and all \purple noise sources (cf. Eq.~(\ref{eq:est_noise_cov}) and Table \ref{tab:noise}), $L=128$  for the \blue, \black, all \yellow, and all \pink noise sources, and $L=512$ for the \green noise source. A ready-to-use implementation of this method is available in the open-source \texttt{Julia} package~\cite{NoiseModels_jl}. Once estimated, these parameters are used to generate synthetic noise samples of arbitrary duration. We illustrate the results for the \blue channel (Gaussian noise) and the \red channel (impulsive noise) in Figure~\ref{fig:noise_panel}. 

For the \blue channel, the power spectral density (PSD)  of the generated  noise and the noise recorded  on the top array element are compared in Figure~\ref{fig:noise_a}, showing close agreement across the signal band. Figure~\ref{fig:noise_b} presents a Q-Q plot comparing the recorded and generated noise samples, where the different curves correspond to different hydrophones. The solid line with unit slope through the origin indicates that the two noises share the same distribution. Time-domain samples of the recorded and generated noises on the top array element are shown in Figure~\ref{fig:noise_d}.

For the \red channel, we compute the pseudo PSD, illustrated in Figure~\ref{fig:noise_e} for the generated noise and the noise recorded  on the top array element. Since the second-order moments are undefined for $\alpha$-stable distributions  that characterise this noise with $\alpha<2$,  the PSD of the \red noise is not well defined, and the pseudo PSD is computed instead as  the Fourier transform of a lower-order statistic \cite{Chitre2025}.
The two spectra exhibit close agreement within the signal band. Figure~\ref{fig:noise_f} shows the relevant Q-Q plots,  confirming agreement between the two noise distributions.
The time-domain samples of the recorded and generated noises are illustrated in Figure~\ref{fig:noise_h}, showing that the impulsive properties of the noise are preserved. 

\begin{figure*}[h]
\centering
    \begin{subfigure}[b]{0.45\textwidth}
        \includegraphics[width=1\columnwidth]{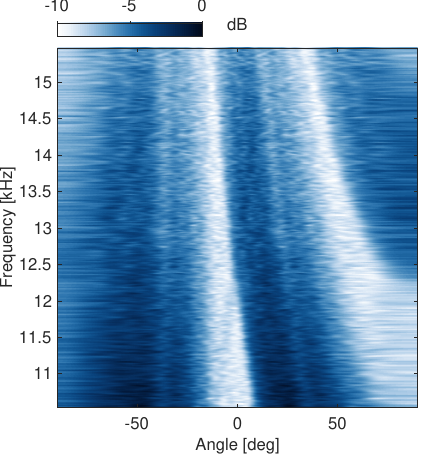}
        \caption{At-sea recording.} \label{fig:freq_angle_recorded}
    \end{subfigure}
    \hfill
    \begin{subfigure}[b]{0.45\textwidth}
        \includegraphics[width=1\textwidth]{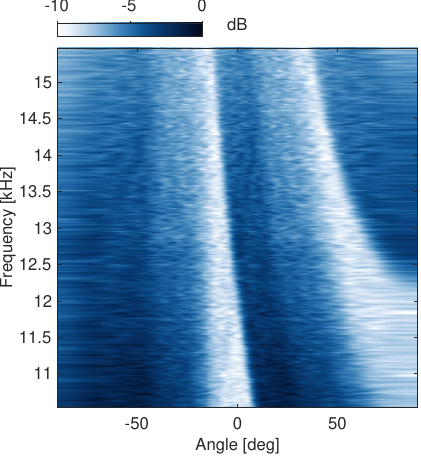}
        \caption{Generated.} \label{fig:freq_angle_generated}
    \end{subfigure}
\caption{Frequency-angle power distribution of the recorded (a) and generated (b) \blue noise. The close agreement between the two confirms that the estimation-generation method preserves the spatial structure of the noise across the array.}
\label{fig:noise_panel_2}
\end{figure*}

Finally, to verify that the spatial correlation properties are also preserved, we estimate the frequency--angle power distribution of the noise by beamforming non-overlapping blocks of the multichannel data with a  steering vector's  angles ranging between $-90^\circ$ and $90^\circ$ relative to the broadside of the vertical array. Figure~\ref{fig:noise_panel_2} shows the resulting frequency-angle power distributions for the recorded (left) and generated (right) \blue noises, respectively, confirming that the generated noise reproduces the spatial structure of the original data.

\section{Verification}
\label{sec:verif}

To verify that the channel library produces realistic simulation results, we compare the performance of a multichannel  \ac{dfe}~\cite{stojanovic_phase-coherent_1994} operating on at-sea recordings with the same receiver operating on signals replayed through the extracted channel. The receiver combines adaptive equalization and phase tracking, and the output \ac{snr} serves as the performance metric.

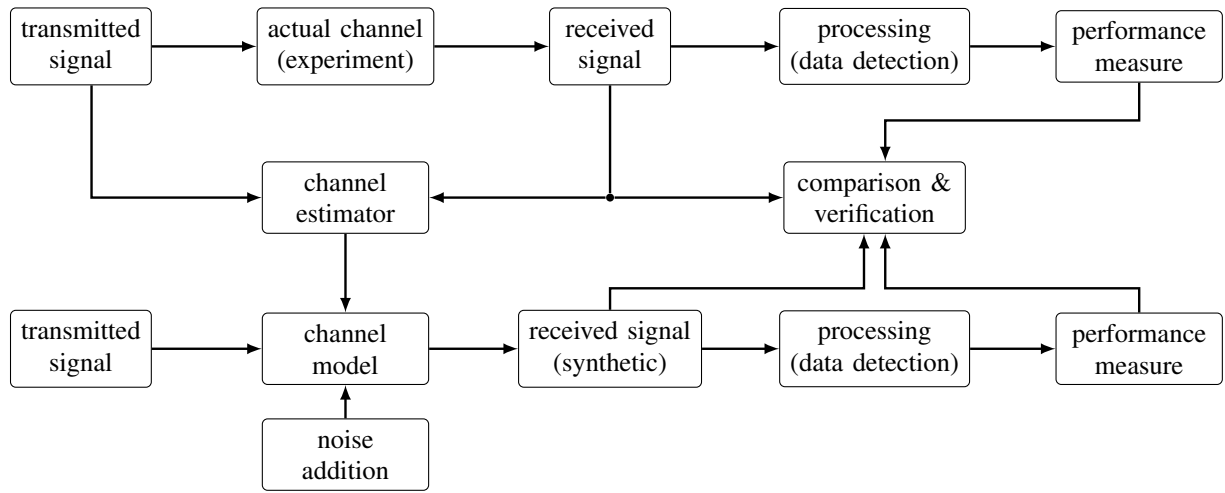
\begin{figure*}[ht]
    \centering
    \tikzset{>=latex}

\begin{tikzpicture}[
    block/.style = {draw, rectangle, rounded corners=2pt, minimum width=2.2cm,
                    minimum height=0.9cm, align=center, font=\small},
    lbl/.style   = {draw, rectangle, rounded corners=2pt, minimum width=1.6cm,
                    minimum height=0.9cm, align=center, font=\small},
    arrow/.style = {->, thick, line width=0.9pt},
    dot/.style   = {circle, fill=black, inner sep=0pt, minimum size=3pt},
]

\def\colA{0}
\def\colB{3.5}
\def\colC{7}
\def\colD{10.5}
\def\colE{14}

\def\rowT{0}
\def\rowM{-2}
\def\rowB{-4}

\def\rowBM{-3.2}

\node[lbl]   (tx1)   at (\colA, \rowT) {transmitted\\signal};
\node[block] (ch1)   at (\colB, \rowT) {actual channel\\(experiment)};
\node[lbl]   (rx1)   at (\colC, \rowT) {received\\signal};
\node[block] (proc1) at (\colD, \rowT) {processing\\(data detection)};
\node[block] (perf1) at (\colE, \rowT) {performance\\measure};

\draw[arrow] (tx1)   -- (ch1);
\draw[arrow] (ch1)   -- (rx1);
\draw[arrow] (rx1)   -- (proc1);
\draw[arrow] (proc1) -- (perf1);

\node[lbl]   (tx2)   at (\colA, \rowB) {transmitted\\signal};
\node[block] (ch2)   at (\colB, \rowB) {channel\\model};
\node[lbl]   (rx2)   at (\colC, \rowB) {received signal\\(synthetic)};
\node[block] (proc2) at (\colD, \rowB) {processing\\(data detection)};
\node[block] (perf2) at (\colE, \rowB) {performance\\measure};

\draw[arrow] (tx2)   -- (ch2);
\draw[arrow] (ch2)   -- (rx2);
\draw[arrow] (rx2)   -- (proc2);
\draw[arrow] (proc2) -- (perf2);

\node[block] (est)  at (\colB, \rowM) {channel\\estimator};
\node[block, minimum width=2.4cm] (comp) at (\colD, \rowM) {comparison \&\\verification};

\node[block] (noise) at (\colB, -5.4) {noise\\addition};
\draw[arrow] (noise) -- (ch2);

\coordinate (tx1_corner) at (\colA, \rowM);
\draw[arrow] ([xshift=4pt]tx1.south) -- ([xshift=4pt]tx1_corner) -- (est.west);

\node[dot] (rx1_junc) at (\colC, \rowM) {};
\draw[arrow] (rx1.south) -- (rx1_junc) -- (est.east);

\draw[arrow] (est.south) -- (ch2.north);

\draw[arrow] (rx1_junc) -- (comp.west);

\draw[arrow] (perf1.south) -- ++(0, -0.5) -| ([xshift=4pt]comp.north);

\draw[arrow] (rx2.north) -- (\colC, \rowBM) -| ([xshift=-4pt]comp.south);

\draw[arrow] (perf2.north) -- (\colE, \rowBM) -| ([xshift=4pt]comp.south);

\end{tikzpicture}
    \caption{Workflow diagram of verifying the accuracy of the model.}
    \label{fig:verif}
\end{figure*}

Figure~\ref{fig:verif} gives a block diagram of the validation procedure. The signals entering the actual channel are the probing waveforms from which the channel estimates have been extracted, whereas the replay signals use the same waveform parameters but a different pseudo-random symbol sequence. Site-specific synthetic noise is added in accordance with the models of Section~\ref{sec:noise} to match the at-sea SNR. The receiver is operated in training mode and we consider  both single-hydrophone  reception   and multi-hydrophone reception with  array processing  applied to all the hydrophones of the array. For the {\green} channel, there is no hydrophone array; however, we exploit temporal diversity in this case  by feeding the receiver with multiple copies of the signal received in different time slots. Note that when generating the noise for multi-slot operation, the noise generator should be invoked repeatedly  to produce independent realizations across slots.

The output \ac{snr} is computed after convergence of the adaptive filter as
\begin{equation}
	\text{SNR}_{\text{out}} = \frac{\sum_{n} |d(n)|^2}{\sum_{n} |d(n) - \hat{d}(n)/\gamma|^2}
	\label{eq:snr_out}
\end{equation}
where  $d(n)$ are the known transmitted data symbols, with $|d(n)|=1$, and $\hat{d}(n)$ are the estimated symbol values. A bias correction factor $\gamma = \mathbb{E}[\hat{d}(n) d^*(n)]$ is applied to account for any residual scaling in the equalizer output.

\begin{figure*}[h]
\centering
\pgfdeclareplotmark{hexagram}{%
    \pgfpathmoveto{\pgfqpointpolar{90}{4pt}}%
    \pgfpathlineto{\pgfqpointpolar{210}{4pt}}%
    \pgfpathlineto{\pgfqpointpolar{330}{4pt}}%
    \pgfpathclose\pgfusepath{fill,stroke}%
    \pgfpathmoveto{\pgfqpointpolar{30}{4pt}}%
    \pgfpathlineto{\pgfqpointpolar{150}{4pt}}%
    \pgfpathlineto{\pgfqpointpolar{270}{4pt}}%
    \pgfpathclose\pgfusepath{fill,stroke}%
}
\begin{subfigure}[t]{0.48\textwidth}
\centering
\begin{tikzpicture}
\begin{axis}[
    width=0.95\linewidth,
    height=0.95\linewidth,
    xlabel={Output SNR at sea [dB]},
    ylabel={Replay output SNR [dB]},
    ylabel style={at={(-0.08,0.5)}},
    xmin=-12, xmax=24,
    ymin=-12, ymax=24,
    xtick={-12,-8,...,24},
    ytick={-12,-8,...,24},
    grid=both,
    label style={font=\small},
    tick label style={font=\small},
    legend style={at={(0.97,0.03)}, anchor=south east, font=\small},
    legend cell align=left,
    clip=false,
]

\addplot[black, no markers, forget plot] coordinates {(-12,-12) (24,24)};

\addplot[gray, dashdotted, no markers, forget plot] coordinates {(-12,-9) (21,24)};
\addplot[gray, dashdotted, no markers, forget plot] coordinates {(-9,-12) (24,21)};

\node[gray, font=\tiny, rotate=45, fill=white, inner sep=1.5pt] at (axis cs:12,15) {$+3$~dB};
\node[gray, font=\tiny, rotate=45, fill=white, inner sep=1.5pt] at (axis cs:15,12) {$-3$~dB};

\addplot[only marks, mark=hexagram, mark size=4pt,
    color=black, mark options={fill=black}]
    coordinates {(23.14, 22.89)};
\addlegendentry{\textcolor{black}{\texttt{Black}}}

\addplot[only marks, mark=x, mark size=5pt,
    color=blue, line width=1.5pt]
    coordinates {(10.28, 10.52)};
\addlegendentry{\textcolor{blue}{\texttt{Blue\_6}}}

\addplot[only marks, mark=diamond*, mark size=4pt,
    color=brown, mark options={fill=brown}]
    coordinates {(-8.72, -9.10)};
\addlegendentry{\textcolor{brown}{\texttt{Brown}}}

\addplot[only marks, mark=triangle*, mark size=4pt,
    color=green, mark options={fill=green, rotate=180}]
    coordinates {(-1.35, -0.19)};
\addlegendentry{\textcolor{green}{\texttt{Green}}}

\addplot[only marks, mark=triangle*, mark size=4pt,
    color=pink, mark options={fill=pink, rotate=-90}]
    coordinates {(9.20, 9.55)};
\addlegendentry{\textcolor{pink}{\texttt{Pink\_1}}}

\addplot[only marks, mark=square*, mark size=3.5pt,
    color=purple, mark options={fill=purple}]
    coordinates {(3.42, 4.20)};
\addlegendentry{\textcolor{purple}{\texttt{Purple\_2}}}

\addplot[only marks, mark=*, mark size=3.5pt,
    color=red, mark options={fill=red}]
    coordinates {(3.01, 4.49)};
\addlegendentry{\textcolor{red}{\texttt{Red\_1}}}

\addplot[only marks, mark=triangle*, mark size=4pt,
    color=yellow, mark options={fill=yellow}]
    coordinates {(0.81, 2.94)};
\addlegendentry{\textcolor{yellow}{\texttt{Yellow\_2}}}

\end{axis}
\end{tikzpicture}
\caption{Single hydrophone}
\label{fig:output_snr_single}
\end{subfigure}
\hfill
\begin{subfigure}[t]{0.48\textwidth}
\centering
\begin{tikzpicture}
\begin{axis}[
    width=0.95\linewidth,
    height=0.95\linewidth,
    xlabel={Output SNR at sea [dB]},
    ylabel={Replay output SNR [dB]},
    ylabel style={at={(-0.08,0.5)}},
    xmin=0, xmax=36,
    ymin=0, ymax=36,
    xtick={0,4,...,36},
    ytick={0,4,...,36},
    grid=both,
    label style={font=\small},
    tick label style={font=\small},
    legend style={at={(0.97,0.03)}, anchor=south east, font=\small},
    legend cell align=left,
    clip=false,
]

\addplot[black, no markers, forget plot] coordinates {(0,0) (36,36)};

\addplot[gray, dashdotted, no markers, forget plot] coordinates {(0,3) (33,36)};
\addplot[gray, dashdotted, no markers, forget plot] coordinates {(3,0) (36,33)};

\node[gray, font=\tiny, rotate=45, fill=white, inner sep=1.5pt] at (axis cs:19,22) {$+3$~dB};
\node[gray, font=\tiny, rotate=45, fill=white, inner sep=1.5pt] at (axis cs:22,19) {$-3$~dB};

\addplot[only marks, mark=hexagram, mark size=4pt,
    color=black, mark options={fill=black}]
    coordinates {(29.84, 30.58)};
\addlegendentry{\textcolor{black}{\texttt{Black}}}

\addplot[only marks, mark=x, mark size=5pt,
    color=blue, line width=1.5pt]
    coordinates {(15.19, 16.96)};
\addlegendentry{\textcolor{blue}{\texttt{Blue\_6}}}

\addplot[only marks, mark=diamond*, mark size=4pt,
    color=brown, mark options={fill=brown}]
    coordinates {(5.41, 3.01)};
\addlegendentry{\textcolor{brown}{\texttt{Brown}}}

\addplot[only marks, mark=triangle*, mark size=4pt,
    color=green, mark options={fill=green, rotate=180}]
    coordinates {(15.81, 19.57)};
\addlegendentry{\textcolor{green}{\texttt{Green}}}

\addplot[only marks, mark=triangle*, mark size=4pt,
    color=pink, mark options={fill=pink, rotate=-90}]
    coordinates {(23.31, 23.47)};
\addlegendentry{\textcolor{pink}{\texttt{Pink\_1}}}

\addplot[only marks, mark=square*, mark size=3.5pt,
    color=purple, mark options={fill=purple}]
    coordinates {(18.18, 20.10)};
\addlegendentry{\textcolor{purple}{\texttt{Purple\_2}}}

\addplot[only marks, mark=*, mark size=3.5pt,
    color=red, mark options={fill=red}]
    coordinates {(8.79, 10.57)};
\addlegendentry{\textcolor{red}{\texttt{Red\_1}}}

\addplot[only marks, mark=triangle*, mark size=4pt,
    color=yellow, mark options={fill=yellow}]
    coordinates {(8.53, 10.36)};
\addlegendentry{\textcolor{yellow}{\texttt{Yellow\_2}}}

\end{axis}
\end{tikzpicture}
\caption{Array processing}
\label{fig:output_snr_array}
\end{subfigure}
\caption{Output SNR obtained after processing the signals that passed through the library channel models (replay) versus the output SNR obtained after processing the actual at-sea recorded signals. Points closer to the diagonal indicate better agreement between  replay and at-sea performance. The dash-dotted gray lines indicate $\pm 3$~dB offsets from the diagonal. The figure on the left shows the comparison when a single receiving element is used and the figure on the right gives the performance comparison for array processing.}
\label{fig:output_snr}
\end{figure*}
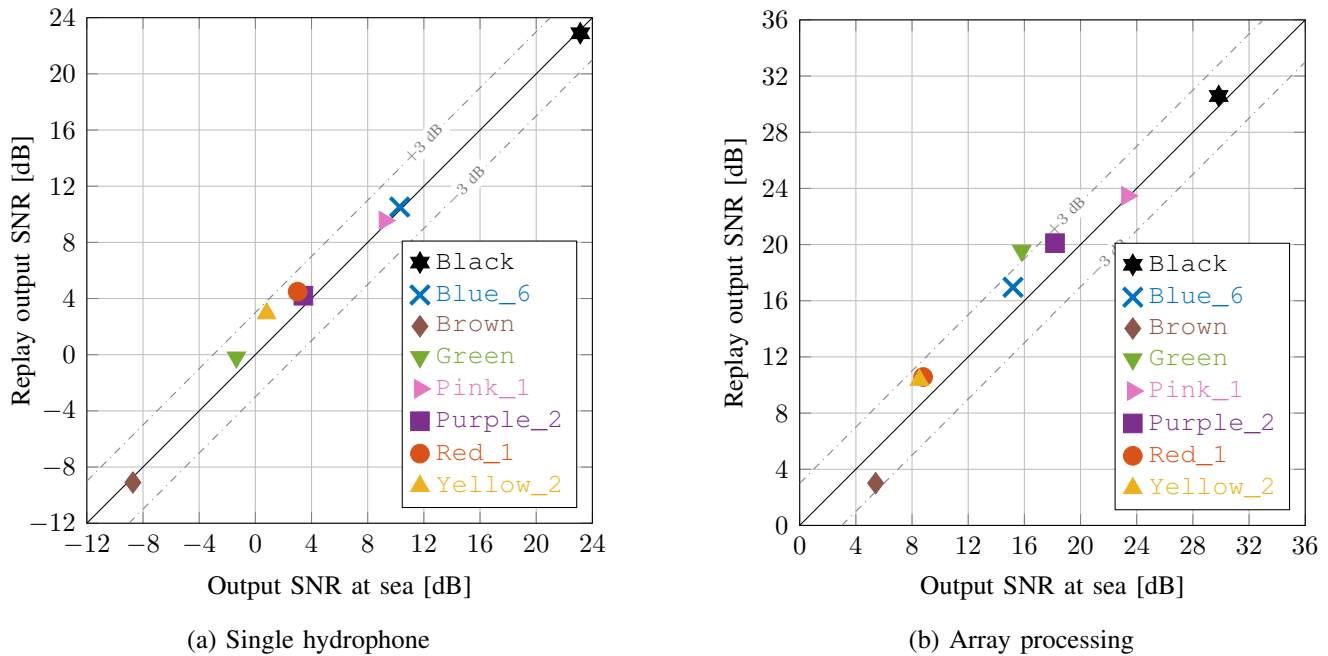

Figure~\ref{fig:output_snr} shows the results of performance validation. The replay \acp{snr} are generally slightly higher than the at-sea values, indicating that the replay channels are more benign for the  modulation scheme used. However, the differences are small, averaging at 1~dB across the set for the single-hydrophone scenario and 2~dB for array processing.

Receiver parameters were coarsely adjusted to each channel, without fine-tuning to optimize the performance for each case. This is of no concern for validation, as what  matters here  is that the \emph{same} receiver algorithm is applied to the at-sea and the replay signals, with the \emph{same} parameter settings. The validation results  demonstrate not only a good agreement, but also  that the repository provides access to communication channels with a wide range of complexities and performance levels.

\begin{figure*}[h]
\centering
\begin{subfigure}[t]{0.45\textwidth}
    \centering
    \includegraphics[width=\linewidth]{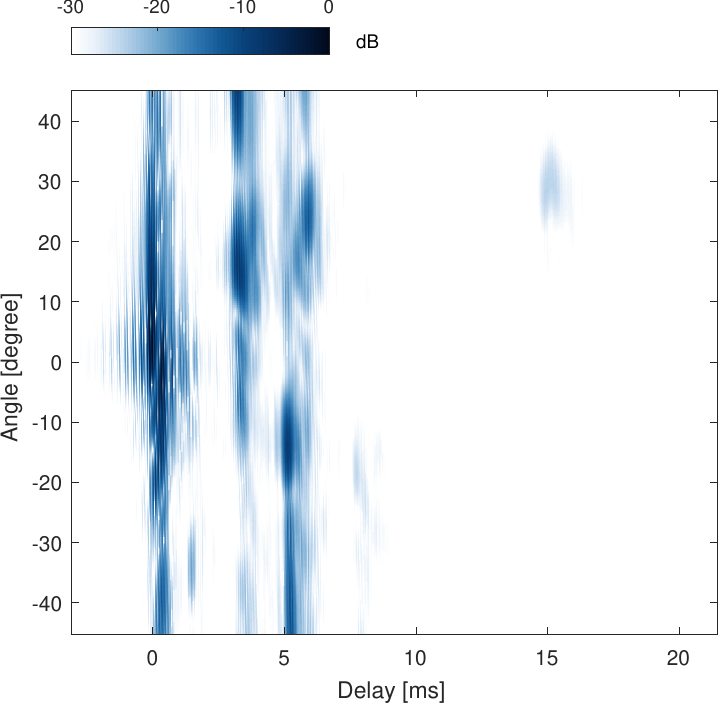}
    \caption{At-sea recording}
    \label{fig:delay_angle_raw}
\end{subfigure}
\hfill
\begin{subfigure}[t]{0.45\textwidth}
    \centering
    \includegraphics[width=\linewidth]{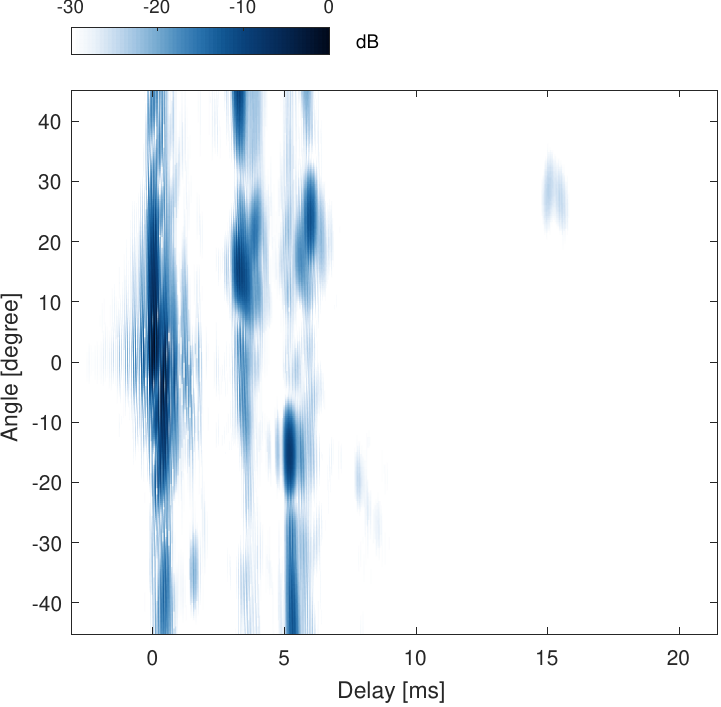}
    \caption{Replay}
    \label{fig:delay_angle_replay}
\end{subfigure}
\caption{Delay-angle spectrum for the  \purple channel with  vertical array, estimated from the at-sea recording (left) and the replayed signal (right). The close agreement in arrival angles and relative intensities confirms that the replay preserves the inter-element delay structure of the vertical array.}
\label{fig:delay_angle}
\end{figure*}

To further assess the  fidelity of the replay model, Figure~\ref{fig:delay_angle} compares the delay-angle spectrum estimated from the at-sea recording with that obtained from the replayed signal. The delay-angle spectrum is computed using a broadband beamformer~\cite{cuji2026path} applied to the signals received  across the vertical array. The two plots show close agreement in both the angular locations and relative intensities of the dominant arrivals, confirming that  replay preserves the inter-element delay structure of the array. This feature  is important for applications that rely on spatial signal processing, as it indicates that the extracted channel faithfully captures the differential propagation delays across the array hydrophones.

\section{Using the Library}
\label{sec:use}

The channel library provides two primary functions for working with the stored channel data: \texttt{replay}, which passes a user-designed signal through the measured channel, and \texttt{unpack}, which reconstructs the full time-varying impulse response. Additionally, the \texttt{noisegen} function generates synthetic  underwater acoustic noise based on statistics collected during the experiments. This section illustrates the usage of the library  functions. The library currently supports \texttt{MATLAB}, \texttt{Python}, and \texttt{Julia}.

\subsection{Channel Replay}

The channel replay function accepts an arbitrary user-designed passband signal and produces the output signal as if it had been transmitted through the measured underwater acoustic channel. During this process,   the full time-varying channel is reconstructed from the stored compressed representation and applied  to the input signal. Note that while the user-supplied input signal can be of an arbitrary type (modulation, coding, etc.), its bandwidth has to conform to the one used in the actual sea trial in order to provide a meaningful and realistic output. 

Let $s_{\text{in}}(t)$ denote the user's input passband signal sampled at the rate $f_{\text{in}}$. This signal is first brought to baseband using the carrier frequency $f_c$ from Table~\ref{tab:params}. If the signal bandwidth is smaller than the one specified by the channel, the signal can have a different user-defined center frequency, provided that the resulting  band is contained within the specified limits. Energy outside the band supported by the channel will be lost in the replay. 

The baseband signal is given by
\begin{equation}
	u_{\text{in}}(t) = \mathcal{LP} \left[ s_{\text{in}}(t) e^{-j2\pi f_c t} \right]
\end{equation}
where $\mathcal{LP}[\cdot]$ represents the lowpass filtering operation.
The baseband signal is then resampled to match the channel's delay-domain sampling rate $f_s$.

The stored channel estimates $\hat{\underline{\bf h}}[n]$ are sampled in time at a compressed rate, which is typically much lower than $f_s$. To apply the channel at the signal sampling rate, the channel impulse response is first decompressed from the compressed time grid to the signal time grid yielding the  drift-free channel response $\hat{\underline{h}}(\tau, t)$. The time-varying convolution is now performed as
\begin{align}
	\bar{y}_\text{r}(nT_s) &= \sum_k \hat{\underline{h}}(kT_s,\, nT_s)\, u_{\text{in}}(nT_s - kT_s)\, e^{j\hat{\varphi}(nT_s)} \notag \\
	&= \underline{\hat{\mathbf{h}}}^\top[n]\,\mathbf{u}_{\text{in}}(n)\, e^{j\hat{\varphi}(nT_s)}
	\label{eq:replay_conv}
\end{align}
where $\hat\varphi(nT_s)$ is the phase estimate which is stored in the library at the original sampling rate $f_s$.

The final step reintroduces the delay drift that was removed during channel extraction. Recalling Eq.~\eqref{eq:a1}, the replayed signal in baseband, not including noise, is given by
\begin{equation}
	\bar v_{\text{r}}(nT_s) = \mathcal{I}\left[\bar y_\text{r} \left(nT_s + \frac{\hat{\varphi}(nT_s)}{2 \pi f_c} \right)\right]
	\label{eq:drift_reinsertion}
\end{equation}
where $\mathcal{I}[\cdot]$ denotes interpolation. Specifically, we use spline interpolation.

In  cases when there is no delay tracking, the time-varying convolution is performed using the basic channel estimate and the corresponding phase, 
\begin{align}
	\bar{v}_\text{r}(nT_s) &= \sum_k \hat{{h}}(kT_s,\, nT_s)\, u_{\text{in}}(nT_s - kT_s)\, e^{j\hat{\theta}(nT_s)} \notag \\
	&= {\hat{\mathbf{h}}}^\top[n]\,\mathbf{u}_{\text{in}}(n)\, e^{j\hat{\theta}(nT_s)}
	\label{eq:replay_conv_2}
\end{align}

The baseband output is finally resampled to the original input sampling rate $f_{\text{in}}$ and up-converted to passband to yield 
\begin{equation}
	\bar r_{\text{out}}(t) =  \Re\left\{\bar v_{\text{r}}(t) e^{j2\pi f_c t}\right\} 
	\label{eq:passband_output}
\end{equation}
Recall that the stored channel responses are normalized so as to preserve any difference in power across the array elements, as well as across time. Hence, a unit-power input signal spanning the entire duration of the channel record will yield a unit output power per element, on average. The total power equals the number of array elements.

The signal (\ref{eq:passband_output}) represents the noiseless channel output. In the next section, we describe generation of the  noise $\hat n(t)$ that will eventually be added to the signal $\bar r_{\text{out}}(t)$. The noise $\hat n(t)$ is normalized similarly as the signal, to preserve any natural difference in power across the array and  yield the total power equal to the total number of array elements,
and  added to the signal as 
\begin{equation}
    r_{\text{out}}(t)=\bar r_{\text{out}}(t) +\sigma_n \hat n(t)
\end{equation}
where the noise level $\sigma_n$ can be adjusted to obtain a desired \ac{snr}. Note that in the case of impulsive noise (\red channel), the variance is not well defined. However, the scale parameter $c$ in $\alpha$-stable distributions plays a similar role as standard deviation in Gaussian distributions. The noise pseudo-power is thus  defined as $2c^2$~\cite{chitre2007vi}, which reduces to the usual definition of noise power when $\alpha=2$. Normalization of $\hat n(t)$ is performed such that the sum of the noise pseudo-powers across all array elements equals the number of array elements.

For multichannel replay, the procedure described by Eqs.~\eqref{eq:replay_conv}--\eqref{eq:passband_output} is applied independently to each receiving element using its own stored channel estimate and phase trajectory. Because the channel extraction stage synchronizes all elements to the first, the inter-element delay structure is inherently preserved in the replayed output. A user can then generate the noise and add it to the replayed output, as described in the next section. 

The replay function thus accepts a passband input signal and produces the corresponding passband received signal as if it had been transmitted through the measured underwater acoustic channel. The function is invoked as
\begin{verbatim}
  y = replay(input, fs, array_index, channel, start);
\end{verbatim}
where \texttt{input} is the passband signal vector, \texttt{fs} is its sampling frequency, \texttt{array\_index} specifies which hydrophone elements to use, and \texttt{channel} is the structure containing the stored channel estimates. The output \texttt{y} is a matrix with  each column corresponding to one array element and containing the corresponding time-domain signal samples taken at the same sampling rate as the input signal.  An optional \texttt{start} argument specifies the starting time index within the stored channel trace; if omitted, a random starting time is selected on each call. Choosing a different starting time on multiple calls  allows a user to pass the signal through  different time segments of the channel. Note that the total number of different non-overlapping segments is limited by the length of the channel record as well as by the length of the user's signal, and  their starting times need to be chosen accordingly. 

\subsection{Noise Generation}

The \texttt{noisegen} function generates synthetic underwater acoustic noise and supports three modes of operation, which we outline below.

\subsubsection{Generic colored Gaussian noise}

When called with only the signal size and sampling rate, the function 
\begin{verbatim}
  w = noisegen(size(y), fs);
\end{verbatim}
generates spatially uncorrelated  Gaussian noise with a power spectral density that decays at 17~dB per decade, a commonly used model for the ambient ocean noise. 

\subsubsection{Spatially correlated colored Gaussian noise}

When provided with experimentally measured noise statistics, the function
\begin{verbatim}
  w = noisegen(size(y), fs, array_index, noise);
\end{verbatim}
generates colored Gaussian noise with the spatiotemporal correlation structure observed during the channel measurements. The parameters \texttt{fs} and \texttt{array\_index} are defined as before, while the \texttt{noise} structure contains estimated $\beta$ coefficients and related noise parameters. The generated noise samples are obtained by filtering independent standard Gaussian sequences through the mixing coefficients $\beta_{ij}(kT_s)$ as described in Eq.~\eqref{eq:mixing}, with $\alpha = 2$.

\subsubsection{Colored impulsive noise}

When the \texttt{noise} structure contains a characteristic exponent $\alpha < 2$, the same calling syntax
\begin{verbatim}
  w = noisegen(size(y), fs, array_index, noise);
\end{verbatim}
generates colored impulsive noise from a symmetric $\alpha$-stable distribution with the spatiotemporal correlation structure captured by the mixing coefficients. The function automatically selects between the Gaussian and impulsive generators based on the value of the $\alpha$ parameter in the \texttt{noise} structure.

\subsection{Channel Unpacking}

For users who require direct access to the time-varying impulse response, for example, for visualization or custom signal processing, the \texttt{unpack} function reconstructs the full channel from the compressed representation. This function is invoked as 
\begin{verbatim}
  h = unpack(fs_time, array_index, channel);
\end{verbatim}
where \texttt{fs\_time} is the desired sampling rate along the time axis and \texttt{array\_index} specifies the hydrophone elements. The output is a three-dimensional array of size $K \times M \times N_t$, where $K$ is the number of delay taps, $M$ is the number of array elements, and $N_t$ is the number of time snapshots.

The unpacking procedure depends on whether delay tracking was used during channel extraction. When delay tracking is enabled, the stored representation consists of a drift-free impulse response $\hat{\underline{h}}(\tau, t)$ and a phase $\hat\varphi(nT_s)$. The unpacking proceeds as follows:
\begin{enumerate}
    \item Resample the compressed channel estimates from the storage
          rate to the desired time-axis rate \texttt{fs\_time}.
    \item Reintroduce the residual phase by multiplying with
          $e^{j\hat\varphi(nT_s)}$.
    \item Reinsert the delay drift via interpolation as described in
          Eq.~\eqref{eq:drift_reinsertion}.
\end{enumerate}
When delay tracking is not enabled, the stored representation contains the basic channel estimate and the phase  $\hat\theta(nT_s)$. In this case, the unpacking resamples the channel estimates to the desired rate, reintroduces the phase by multiplying with $e^{j\hat\theta(nT_s)}$, and no drift reinsertion is required.

The result of the channel unpacking procedure is the impulse response $\hat h(\tau, t)$, which contains any drift that was present in the channel. Its magnitude can be visualized as a heat map with delay on one axis and time on the other, as in Figure~\ref{fig:slanted}, revealing the temporal evolution of the multipath structure. 

\section{Gallery of Channels}
\label{sec:gal}

\begin{figure*}
    \centering
    \begin{subfigure}[b]{0.24\textwidth}
        \centering
        \includegraphics[width=\textwidth]{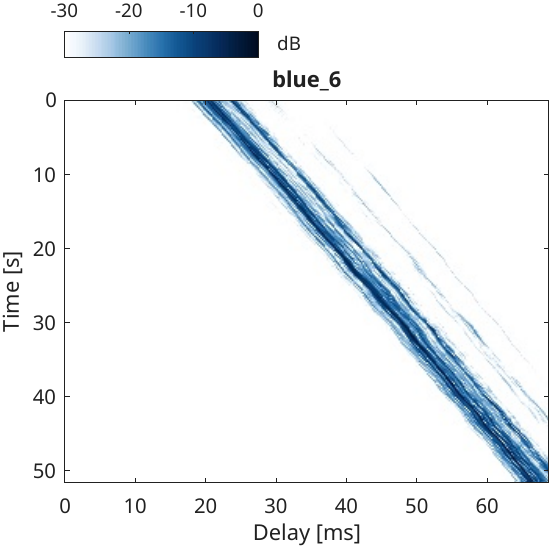}
        \caption{\blue}
    \end{subfigure}
    \hfill
    \begin{subfigure}[b]{0.24\textwidth}
        \centering
        \includegraphics[width=\textwidth]{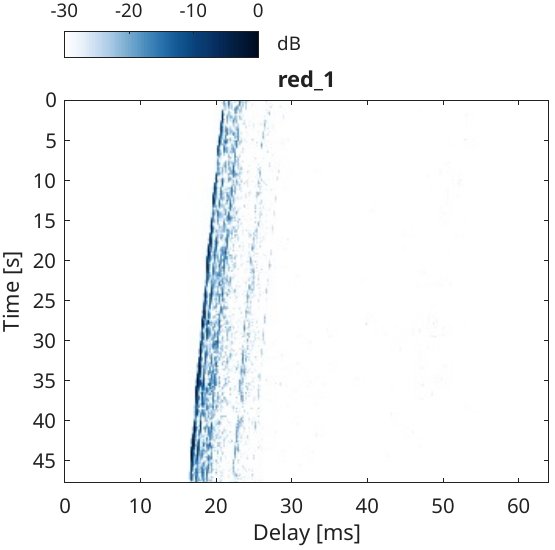}
        \caption{\red}
    \end{subfigure}
    \hfill
    \begin{subfigure}[b]{0.24\textwidth}
        \centering
        \includegraphics[width=\textwidth]{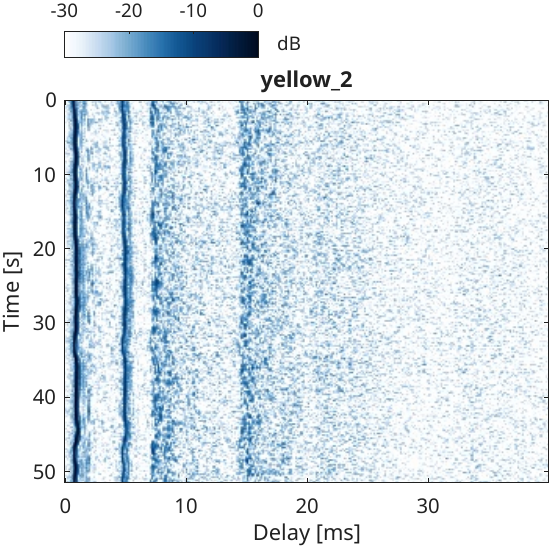}
        \caption{\yellow}
    \end{subfigure}
    \hfill
    \begin{subfigure}[b]{0.24\textwidth}
        \centering
        \includegraphics[width=\textwidth]{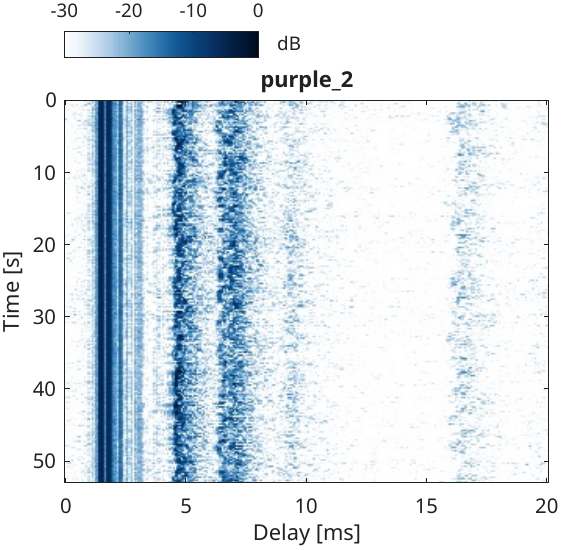}
        \caption{\purple}
    \end{subfigure}

    \vspace{0.5cm}

    \begin{subfigure}[b]{0.24\textwidth}
        \centering
        \includegraphics[width=\textwidth]{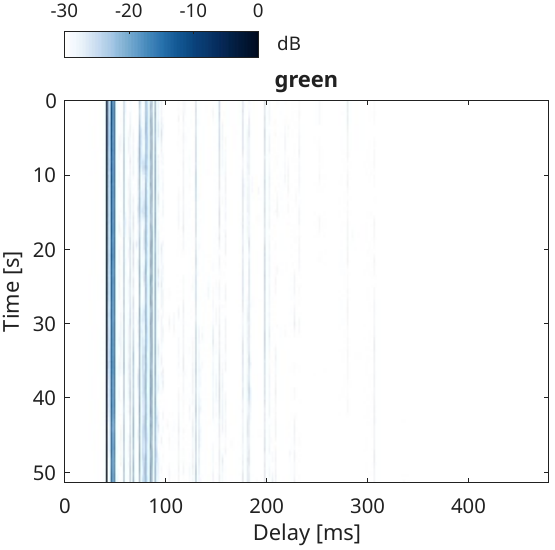}
        \caption{\green}
    \end{subfigure}
    \hfill
    \begin{subfigure}[b]{0.24\textwidth}
        \centering
        \includegraphics[width=\textwidth]{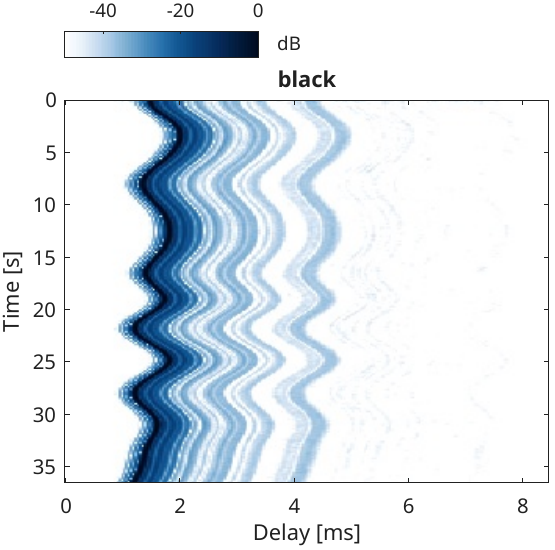}
        \caption{\black}
    \end{subfigure}
    \hfill
    \begin{subfigure}[b]{0.24\textwidth}
        \centering
        \includegraphics[width=\textwidth]{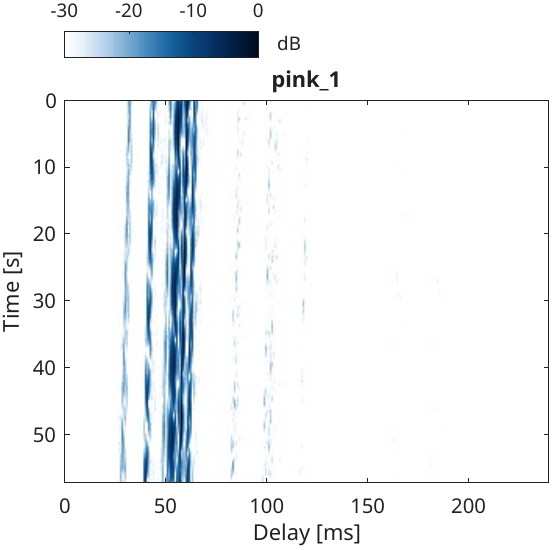}
        \caption{\pink}
    \end{subfigure}
    \hfill
    \begin{subfigure}[b]{0.24\textwidth}
        \centering
        \includegraphics[width=\textwidth]{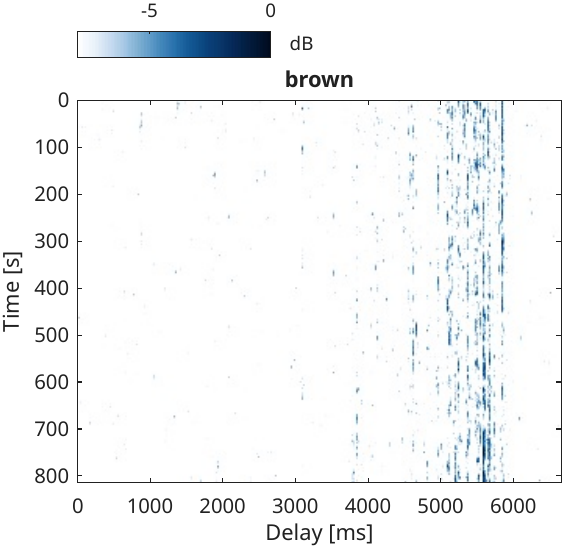}
        \caption{\brown}
    \end{subfigure}

    \caption{Gallery of selected channel impulse response examples from a single receiver element in each of eight underwater acoustic environments spanning various geographic conditions: mobile platform (\blue), short-range coastal (\red), wind-dominated mid-range (\yellow, \purple), reverberant fjord (\green), vertical Mariana Trench (\black), mid-range coastal (\pink), and trans-Pacific non-minimum phase (\brown). The channels span frequencies from 75~Hz to 25~kHz, ranges from 60~m to 3250~km, and water depths from 8~m to 8720~m. Specific parameters are detailed in Table~\ref{tab:params}.}

    \label{fig:ir_channels}
\end{figure*}

To illustrate the variety of underwater acoustic channels contained in the repository, Figure~\ref{fig:ir_channels} presents the time-varying impulse response estimates from a single receiver element in each of the eight channel collections. These channels span a remarkable range of environmental conditions and experimental configurations: from the shallow, reverberant waters of Martha's Vineyard (\purple, wind-dominated) to the extreme depths of the Mariana Trench (\black); from short-range coastal deployments in Singapore (\red) and the highly reverberant Norwegian fjord (\green) to trans-Pacific propagation over 3250~km (\brown, exhibiting non-minimum phase characteristics). The  frequencies range from 75~Hz in deep ocean basins to 25~kHz in coastal environments. The impulse responses reveal distinct propagation characteristics shaped by factors such as  water depth, transmission distance, sea surface and bottom conditions, geological conditions, frequency band, and platform mobility. Notable features include rough surface scattering in the \yellow and \purple channels, Doppler effects from transmitter motion in the \blue channel,  reverberation patterns of a fjord environment in the \green channel, and the benign characteristics of a vertical link in the \black channel.

This collection provides researchers with a comprehensive testbed representing the wide spectrum of challenges encountered in underwater acoustic communications, from highly dispersive shallow-water channels to stable deep-water paths, enabling robust algorithm development and performance evaluation across various operational scenarios.

\section{Conclusion}
\label{sec:concl}

We have presented an open-access library of underwater acoustic channels derived from field experiments spanning geographically diverse locations and a wide range of propagation conditions. Each channel in the library consists of time-varying impulse responses extracted from at-sea recordings together with a site-specific ambient noise model, and is accompanied by a replay interface that allows users to pass arbitrary signals through realistic underwater acoustic channels in a reproducible manner. Validation against the original at-sea recordings confirms that the replayed channels faithfully preserve the communication-relevant characteristics of the measured data.

The library is designed to grow. Researchers who have access to field recordings are encouraged to contribute new channels by following the extraction procedures described in this paper. Contributions from different environments and experimental configurations will strengthen the library's role as a common framework for comparing communication, networking, and signal processing algorithms. Guidelines for preparing and submitting new channels are available on the project website~\cite{uwachannels_website}. To facilitate this process, the channel estimation and noise extraction scripts used to build the library are also open-sourced~\cite{uwachannels_github,NoiseModels_jl}, enabling researchers to generate new channel entries directly from their own field recordings.

A natural next step is to complement the replay-based channels with statistical channel models, and standardize these models. While replay offers high fidelity to a specific measurement, statistical models allow unlimited time-series as well as exploration of conditions beyond those captured in any single recording -- for example, by varying parameters such as range, water depth, or sea state. Developing such models from the growing body of data in the library and integrating them into the same interface is an important direction for future work. Our hope is that this effort will serve as a step toward the adoption of standardized underwater acoustic channel models.

\section{Acknowledgments}

This project was supported in part by the Office of Naval Research grant \texttt{N00014-23-1-2852}.  We would like to thank all those who were involved in conducting the experiments  that produced the invaluable signal recordings used in this work, as well as the funding agencies that made the experiments possible and granted permission to publish the findings reported in this article: Office of Naval Research in the USA, SFI Smart Ocean in Norway, Acoustic Research Laboratory at the National University of Singapore, and the Japan Agency for Marine-Earth Science and Technology in Japan. Last but not least, we  are grateful to all our beta testers who provided  invaluable constructive feedback. 

\bibliographystyle{IEEEtran}
\bibliography{ref}

\end{document}